\documentclass[notitlepage,pallis,twoside]{wpallis}
\usepackage{fancyh}
\usepackage[l]{floatflt}
\usepackage{epsfig}
\usepackage{amssymb}
\usepackage{latexsym}
\usepackage{times}
\usepackage{slashed,cite}
\usepackage{upgreek}
\numberwithin{equation}{section}

\newcommand{\ec}{\end{center}}
\newcommand{\bec}{\begin{center}}

\newcommand{\eem}{\end{matrix}}
\newcommand{\bem}{\begin{matrix}}
\newcommand{\eeq}{\end{equation}}
\newcommand{\beq}{\begin{equation}}
\newcommand{\ba}{\begin{array}}
\newcommand{\ea}{\end{array}}
\newcommand{\bea}{\begin{eqnarray}}
\newcommand{\eea}{\end{eqnarray}}
\newcommand{\baq}{\begin{eqnarray}}
\newcommand{\eaq}{\end{eqnarray}}

\newcommand\eqs[2]{Eqs.~(\ref{#1}) and (\ref{#2})}

\newcommand\eqss[3]{Eqs.~(\ref{#1}), (\ref{#2}) and (\ref{#3})}

\newcommand{\ftn}{\footnotesize}

\newcommand{\ssz}{\scriptsize}
\newcommand{\TeV}{{\mbox{\rm TeV}}}

\newcommand{\GeV}{{\mbox{\rm GeV}}}

\newcommand{\sFref}[2]{Fig.~\ref{#1}-{\ftn\sf ({#2})}}
\newcommand{\sEref}[2]{Eq.~(\ref{#1}{\ftn\sf {#2}})}

\newcommand{\etal}{{\it et al.\/}}

\def\lf{\left(}
\def\rg{\right)}

\newcommand{\ck}{\ensuremath{c_\mathcal{R}}}
\newcommand{\ns}{\ensuremath{n_{\rm s}}}
\newcommand{\as}{\ensuremath{\alpha_{\rm s}}}
\newcommand{\Dex}{\ensuremath{\Delta_{\rm m*}}}
\newcommand{\Dcex}{\ensuremath{\Delta_{\rm c*}}}

\newcommand{\rcc}{\ensuremath{\mathcal{R}}}
\newcommand{\rce}{\ensuremath{\widehat{\mathcal{R}}}}
\newcommand{\Ve}{\ensuremath{\widehat{V}}}

\newcommand{\Trh}{\ensuremath{T_{\rm rh}}}
\newcommand\sig{\ensuremath{\langle \sigma \rangle}}
\newcommand{\sgc}{\ensuremath{\sigma_{\rm c}}}

\newcommand{\sg}{\ensuremath{\sigma}}
\newcommand{\ld}{\ensuremath{\lambda}}
\newcommand{\Ld}{\ensuremath{\Lambda}}
\newcommand{\kp}{\ensuremath{\kappa}}
\newcommand{\se}{\ensuremath{\widehat\sigma}}
\newcommand{\sgb}{\ensuremath{\bar\sigma}}
\newcommand{\sex}{\ensuremath{\widehat{\sigma}_*}}

\newcommand{\geu}{\ensuremath{\widehat g}}

\newcommand\Gm[1]{\Gamma_{#1}}

\newcommand{\mP}{\ensuremath{m_{\rm P}}}

\def\trns{trans-Planckian}
\def\sub{sub-Planckian~}

\begin{document}

\thispagestyle{empty}
%%%%%%%%%%%%%%%

\title[]{\Large\bfseries\scshape Non-Minimally Gravity-Coupled Inflationary Models}

\author{\large\bfseries\scshape C. Pallis}
\address[] {\sl Department of Physics, University of Cyprus, \\
P.O. Box 20537, CY-1678 Nicosia, CYPRUS}

% Physics Division, School of Technology, \\ Aristotle
%University of Thessaloniki,{\tt pallis.constantinos@ucy.ac.cy}
%\\ GR-541 24 Thessaloniki,  GREECE\\ {\tt kpallis@auth.gr}

\begin{abstract}{\small {\bfseries\scshape Abstract} \\
\par We consider the non-supersymmetric models of chaotic
(driven by a quadratic potential) and hybrid inflation, taking
into account the minimal possible radiative corrections to the
inflationary potential. We show that two simple coupling functions
$f(\sg)$ (with a parameter $\ck$ involved) between the inflaton
field $\sg$ and the Ricci scalar curvature ensure, for
sub-Planckian values of the inflaton field, observationally
acceptable values for the spectral index, $\ns$, and sufficient
reheating after inflation. In the case of chaotic inflation we
consider two models with large $\ck$'s resulting to
$\ns\simeq0.955$ or $0.967$ and tensor-to-scalar ratio
$r\simeq0.2$ or $0.003$, respectively. In the case of hybrid
inflation, the selected $f(\sg)$ assists us to obtain hilltop-type
inflation. For values of the relevant mass parameter, $m$, less
than $10^6~\TeV$ and the observationally central value of $\ns$,
we find $\ck\simeq(0.015-0.078)$ with the relevant coupling
constants $\lambda=\kappa$ and the symmetry breaking scale, $M$,
confined in the ranges $(2\cdot 10^{-7}-0.001)$ and
$(1-16.8)\cdot10^{17}~\GeV$, respectively.}
%
%\\ \\{\ftn \sc Keywords}: {\ftn Cosmology};
%{\ftn\sc PACS codes:} {\ftn 98.80.Cq}
\end{abstract} \maketitle
\publishedin{{\sl  Phys. Lett. B} {\bf 692}, 287 (2010)}

\thispagestyle{empty}

\setcounter{page}{1} \pagestyle{fancyplain}

%\addtolength{\headheight}{.5cm}

\rhead[\fancyplain{}{ \bf \thepage}]{\fancyplain{}{\sl
Non-Minimally Gravity-Coupled Inflationary Models}}
\lhead[\fancyplain{}{\sl \leftmark}]{\fancyplain{}{\bf \thepage}}
\cfoot{}

\vspace*{-.cm}

\section{Introduction}\label{intro}

\emph{Non-minimal inflation} (non-MI) \cite{wmap3} i.e. inflation
constructed in the presence of a non-minimal coupling between the
inflaton field and the Ricci scalar curvature, $\rcc$, has gained
a fair amount of recent attention \cite{sm1, shw, nmc}. In
particular, it is shown that non-MI can be realized within the
\emph{Standard Model} (SM) -- or minimal extensions \cite{love} of
it -- provided the inflaton couples strongly enough to $\rcc$. The
role of inflaton can be played either by the Higgs doublet either
by a SM singlet coupled to Higgs. Although quite compelling,
non-MI within SM suffers from {\sf (i)} several computational
uncertainties regarding the impact of the quantum corrections in
the presence of such a strong non-minimal gravitational coupling
and {\sf (ii)} the ambiguity about the hierarchy between the
cutoff scale of the effective theory and the energy scale of the
inflationary plateau \cite{cutoff, john}. Be that as it may, it
would be interesting to examine if appropriately selected
non-minimal gravitational couplings can have beneficial
consequences -- as for the reconstruction of the cosmic expansion
history \cite{nojiri} -- for other well-motivated and rather
natural models of inflation (for a survey see, e.g.,
\cref{review}).

Two such models are undoubtedly \emph{Chaotic} (CI) \cite{chaotic}
and \emph{Hybrid Inflation} (HI) \cite{hybrid}. In this paper we
focus on the non-supersymmetric version of these models. CI driven
by a quadratic potential provides the simplest realization of
inflation without initial-value problem and with quite interesting
predictions for the (scalar) spectral index, $\ns$, and the
scalar-to-tensor ratio, $r$. However, trans-Planckian
inflaton-field values are typically required to allow for a
sufficiently long period of inflation. Thus non-renormalizable
corrections from quantum gravity are expected to destroy the
flatness of the potential, invalidating thereby CI. On the other
hand, HI -- although can be accommodated with sub-Planckian values
for the inflaton -- suffers from the problem of the enhanced $\ns$
which turns out to be, mostly, well above the prediction of the
fitting \cite{wmap} of the five-year results from the
\emph{Wilkinson Microwave Anisotropy Probe Satellite} (WMAP5) plus
\emph{baryon-acoustic-oscillations} (BAO) and \emph{supernovae}
(SN) data -- for an up-to-date analysis of the problem of initial
conditions within HI, see \cref{rocher}.

\tableofcontents\vskip-1.3cm\noindent\rule\textwidth{.4pt}\\

Note, in passing, that the introduction of \emph{supersymmetry}
(SUSY) and its local extension -- supergravity (SUGRA) -- can
alleviate the shortcomings of both models -- see \cref{chaotic1}
for several resolutions to the problem of CI and \cref{susyhybrid,
gpp, battye, mhi, hinova} for proposals related to the
disadvantage of HI. However, we have to accept that there is no
direct experimental confirmation of SUSY until now. On the
contrary, there is a strong observational evidence in favor of the
inflationary paradigm. Consequently, it is worthwhile to build
models of CI and HI consistently with the observations, even
without the presence of SUSY -- for similar recent attempts, see
\cref{circ, hirc}.

In this paper, we propose two types of non-minimal coupling
functions $f(\sigma)$  between the inflaton and $\rcc$ which
support a resolution to the aforementioned problems of CI and HI.
After the end of non-MI, both  $f(\sigma)$'s  shrink to unit
assuring thereby, a safe transition to the Einstein gravity in
time. In the case of \emph{non-minimal CI} (non-MCI), two models
with clearly distinctive results are investigated. In the case of
\emph{non-minimal HI} (non-MHI), the inflationary trajectory is
concave downwards and so, inflation turns out to be of
hilltop-type \cite{lofti}. In both cases, the minimal possible
radiative corrections \cite{coleman} to the inflationary potential
are considered, sub-Planckian values of the inflaton field are
required and adequate reheating of the universe is accomplished
via curvature-induced \cite{reheating} couplings of the inflaton
to matter fields. Comparisons with the results obtained for the
minimal version of both inflationary models are also displayed.

Below, we describe the generic formulation of non-MI
(Sec.~\ref{sugra}) and then apply the relevant results, for
appropriate choices of $f(\sg)$, in the case of non-MCI and
non-MHI in Sec.~\ref{ci} and \ref{hi} respectively. Finally,
Sec.~\ref{con} summarizes our conclusions. Throughout the text, we
set natural units for the Planck's constant, Boltzmann's constant
and the velocity of light ($\hbar=c=k_{\rm B}=1$) the subscript
$,\chi$ denotes derivation \emph{with respect to} (w.r.t.) the
field $\chi$ (e.g., $_{,\chi\chi}=d^2/d\chi^2$) and a bar over a
field $\chi$ denotes normalization w.r.t the reduced Planck mass,
$\mP = 2.44\cdot 10^{18}$ GeV, i.e., $\bar\chi=\chi/\mP$. Finally,
we follow the conventions of \cref{kolb} for the quantities
related to the gravitational sector of our set-up.

\section{Inflation with non-Minimal Gravitational Coupling}\label{sugra}

Non-MI, by its definition, can be realized by a scalar field
non-minimally coupled to Ricci scalar curvature. The formulation
of a such theory is described in \Sref{sugra1}. Based on it, we
then derive the inflationary observables and impose observational
constraints in \Sref{obs}.

\subsection{non-Minimally Curvature-Coupled Scalar
Theory}\label{sugra1}

The dynamics of a scalar field $\sigma$ non-minimally coupled to
$\rcc$ through a coupling function $f(\sigma)$ is controlled, in
the Jordan frame, by the following action -- see, e.g.,
\cref{shw}:
\beq \label{action1} {\cal S} = \int d^4 x \sqrt{-g}
\left(-\frac{1}{2} \mP^2f(\sigma)\rcc +\frac{1}{2}g^{\mu\nu}
\partial_\mu \sg\partial_\nu \sg- V \left(\sigma\right)\right),
\eeq
where $g$ is the determinant of the background
Friedmann-Robertson-Walker metric \cite{kolb}. To guarantee the
validity of the ordinary Einstein gravity at low energy, we
require $f(\langle\sg\rangle) =1$, where $\sig$ is the
\emph{vacuum expectation value} (v.e.v) of $\sg$ at the end of
non-MI.

The action in \Eref{action1} can be brought in a simpler form by
performing a conformal transformation \cite{conformal} to the
so-called Einstein frame where the the gravitational sector of our
model becomes minimal. Indeed, if we define the Einstein-frame
metric
\beq\geu_{\mu\nu}=f\,g_{\mu\nu}~~\Rightarrow~~\left\{\bem
%\begin{array}{rl}
\sqrt{-\geu}=f^2\sqrt{-g}\hspace*{0.2cm}\mbox{and}\hspace*{0.2cm}
\geu^{\mu\nu}=g^{\mu\nu}/f, \hfill \cr
\widehat\rcc=\left(\rcc+3\Box\ln f+3g^{\mu\nu} \partial_\mu
f\partial_\nu f/2f^2\right)/f \hfill \cr\eem
%\end{array}
\right.\eeq
-- where $\Box=\lf
-g\rg^{-1/2}\partial_\mu\lf\sqrt{-g}\partial^\mu\rg$ and hat is
used to denote quantities defined in the Einstein frame -- and
introduce the Einstein-frame canonically normalized field, $\se$,
and potential, $\Ve$, defined as follows:
\beq \label{VJe}
\left(\frac{d\se}{d\sigma}\right)^2=J^2=\frac{1}{f}+{3\over2}\mP^2\left({f_{,\sigma}\over
f}\right)^2\hspace*{0.2cm}\mbox{and}\hspace*{0.2cm} \Ve(\se) =
\frac{V\lf\se(\sigma)\rg}{f\lf\se(\sigma)\rg^2},\eeq
the action in \Eref{action1} can be simplified, taking the form
\beq {\cal S}= \!\int d^4 x \sqrt{-\geu}\left(-\frac12 \mP^2
\rce+\frac12\geu^{\mu\nu} \partial_\mu \se\partial_\nu
\se-\Ve\lf\se\rg\right). \label{action} \eeq
Based on the action above, we can proceed readily to the analysis
of non-MI in the Einstein frame using the standard slow-roll
approximation \cite{review, lectures} -- see below. It can be
shown \cite{general} that the results calculated this way are the
same as if we had calculated them with the non-minimally coupled
scalar field in the Jordan frame.

One of the outstanding features of the scalar theories with
non-minimal $f(\sg)$ is that $\sg$ can decay via gravitational
effects \cite{reheating} even without explicit couplings between
$\sg$ and matter fields. This is, because couplings arise
spontaneously when $\sg$ settles in its v.e.v, \sig, and
oscillates, with coupling constants involving derivatives of
$f(\sg)$ calculated for $\sg=\sig$. If we identify $\sg$ as the
inflaton, these couplings can ensure the reheating of the
universe. Assuming the existence of a bosonic field minimally
coupled to gravity, with negligible mass compared to the mass of
$\sg$, $m_\sg=V_{,\sg\sg}(\sig)^{1/2}$, we get \cite{reheating,
rhodos} for the reheat temperature
\begin{equation}
\Trh\simeq\left(\frac{72}{5\pi^2 g_{\rho*}(\Trh)}\right)^{1/4}
\sqrt{\Gamma_\sg \mP}\hspace*{0.2cm}\mbox{where}\hspace*{0.2cm}
\Gamma_\sg\simeq\frac{f_{,\sg}\left(\sig\right)^2m^3_\sg}{128\pi}\left(
1+{3\over2}\mP^2f_{,\sg}\left(\sig\right)^2\right)^{-1}\label{GTrh}
\end{equation}
is the decay rate of $\sg$, in the regime $T\ll m_\sg$ which is
valid in our applications. Clearly, this construction is
applicable if $f_{,\sg}\left(\sig\right)\neq0$ (and this is valid
for the $f(\sg)$'s considered in \Sref{ci} and \ref{hi}). Also,
assuming the particle spectrum of SM, we set $g_{\rho*}=106.75$
for the relativistic degrees of freedom.

\subsection{Inflationary Observables -- Constraints} \label{obs}

Under the assumption that {\sf (i)} the curvature perturbations
generated by $\sigma$ is solely responsible for the observed
curvature perturbation and {\sf (ii)} there is a conventional
cosmological evolution (see below) after inflation, the
inflationary parameters can be restricted imposing the following
requirements:

\paragraph{(a)} The power spectrum $P_{\cal R}$ of the curvature perturbations
generated by $\sigma$ at the pivot scale $k_{*}=0.002/{\rm Mpc}$
is to be confronted with the WMAP5 data~\cite{wmap},
\begin{equation}  \label{Prob}
P^{1/2}_{\cal R}=\: \frac{1}{2\sqrt{3}\, \pi\mP^3} \;
\frac{\Ve(\sex)^{3/2}}{|\Ve_{,\se}(\sex)|}=\frac{|J(\sigma_*)|}{2\sqrt{3}\,
\pi\mP^3} \;
\frac{\Ve(\sigma_*)^{3/2}}{|\Ve_{,\sigma}(\sigma_*)|}\simeq4.91\cdot
10^{-5},
\end{equation}
where $\sigma_*~[\se_*]$ is the value of $\sg~[\se]$ when $k_*$
crosses outside the inflationary horizon.

\paragraph{(b)} The number of e-foldings, $\widehat N_{*}$, that the
scale $k_*$ suffers during FHI is to account for the total number
of e-foldings $\widehat N_{\rm tot}$ required for solving the
horizon and flatness problems of standard big bag cosmology, i.e.,
$\widehat N_{*}=\widehat N_{\rm tot}$. Specifically, we calculate
$\widehat N_{*}$ through the relation
\begin{equation}
\label{Nhi}  \widehat{N}_{*}=\:\frac{1}{m^2_{\rm P}}\;
\int_{\se_{\rm f}}^{\se_{*}}\, d\se\: \frac{\Ve_{}}{\Ve_{,\se}}=
{1\over\mP^2}\int_{\sigma_{\rm f}}^{\sigma_{*}}\, d\sigma\:
J^2\frac{\Ve_{}}{\Ve_{,\sigma}},
\end{equation}
where $\sg_{\rm f}~[\se_{\rm f}]$ is the value of $\sg~[\se]$ at
the end of inflation, which can be found, in the slow-roll
approximation and for the considered in this paper models, from
the condition
$$ {\sf max}\{\widehat\epsilon(\sigma_{\rm
f}),|\widehat\eta(\sigma_{\rm
f})|\}=1,\hspace*{0.2cm}\mbox{where}$$
\beq \label{sr}\widehat\epsilon=
{\mP^2\over2}\left(\frac{\Ve_{,\se}}{\Ve}\right)^2={\mP^2\over2J^2}\left(\frac{\Ve_{,\sigma}}{\Ve}\right)^2
\hspace*{0.15cm}\mbox{and}\hspace*{0.2cm}\widehat\eta= m^2_{\rm
P}~\frac{\Ve_{,\se\se}}{\Ve}={\mP^2\over J^2}\left(
\frac{\Ve_{,\sigma\sigma}}{\Ve_{}}-\frac{\Ve_{,\sigma}}{\Ve}{J_{,\sg}\over
J}\right)\cdot \eeq
The required $\widehat{N}_{\rm tot}$ at $k_*$ can be easily
derived \cite{hinova} consistently with our assumption of a
conventional post-inflationary evolution. In particular, we assume
that inflation is followed successively by the following three
epochs: {\sf (i)} the decaying-inflaton dominated era which lasts
at a reheat temperature $T_{\rm rh}$, {\sf (ii)} a radiation
dominated epoch, with initial temperature $T_{\rm rh}$, which
terminates at the matter-radiation equality, {\sf (iii)} the
matter dominated era until today. In particular, we obtain -- c.f.
\cref{hinova}
\begin{equation}  \label{Ntot}
\widehat{N}_{\rm tot}\simeq22.4+2\ln{V(\sg_{*})^{1/4}\over{1~{\rm
GeV}}}-{4\over 3}\ln{V(\sg_{\rm f})^{1/4}\over{1~{\rm GeV}}}+
{1\over3}\ln {T_{\rm rh}\over{1~{\rm
GeV}}}+{1\over2}\ln{f(\sg_{\rm f})\over f(\sg_*)},
\end{equation}
where the last term emerges \cite{john} from the transition from
the Jordan to Einstein frame. Note that $\widehat R=\sqrt{f}R$
with $R$ being the scale factor of the universe.

\paragraph{(c)} The (scalar) spectral index, $n_{\rm s}$, its
running, $a_{\rm s}$, and the scalar-to-tensor ratio $r$ are to be
consistent with the fitting \cite{wmap} of the WMAP5 plus BAO and
SN data, i.e.,
\begin{equation}\label{nswmap}
\mbox{\ftn\sf (a)}\hspace*{0.2cm} n_{\rm
s}=0.96\pm0.026,\hspace*{0.2cm}\mbox{\ftn\sf (b)}\hspace*{0.2cm}
-0.068\leq a_{\rm s}\leq0.012
\hspace*{0.2cm}\mbox{and}\hspace*{0.2cm}\mbox{\ftn\sf
(c)}\hspace*{0.2cm}r<0.22,
\end{equation}
at 95$\%$ \emph{confidence level} (c.l.). The observable
quantities above can be estimated through the relations:
\beq\label{ns} n_{\rm s}=\: 1-6\widehat\epsilon_*\ +\
2\widehat\eta_*,\hspace*{0.2cm} \alpha_{\rm s}
=\:{2\over3}\left(4\widehat\eta_*^2-(n_{\rm
s}-1)^2\right)-2\widehat\xi_*\hspace*{0.2cm}
\mbox{and}\hspace*{0.2cm} r=16\widehat\epsilon, \eeq
where $\widehat\xi=\mP^4 {\Ve_{,\widehat\sigma}
\Ve_{,\widehat\sigma\widehat\sigma\widehat\sigma}/\Ve_{}^2}=
\mP^2\,\Ve_{,\sigma}\,\widehat\eta_{,\sigma}/\Ve\,J^2+2\widehat\eta\widehat\epsilon$
and the variables with subscript $*$ are evaluated at
$\sigma=\sigma_{*}$. Note, in passing, that the utilized here non
minimal $f(\sg)$'s do not produce \cite{fnl} observationally
interesting non-gaussianity -- for reviews see, e.g., \cref{gaus}.

\paragraph{\ftn\sf (d)} To avoid corrections from quantum gravity, we
impose two additional theoretical constraints on our models --
keeping in mind that $\Ve(\sg_{\rm f})\leq\Ve(\sg_*)$:
\beq \label{subP}\mbox{\ftn\sf (a)}\hspace*{0.2cm}
\Ve(\sg_*)^{1/4}\leq\mP
\hspace*{0.2cm}\mbox{and}\hspace*{0.2cm}\mbox{\ftn\sf
(b)}\hspace*{0.2cm} \sg_*\leq\mP.\eeq
 Although it is argued \cite{circ, Linde} that violation of
\sEref{subP}{b} may not be necessarily fatal, we insist on
imposing this condition in order to deliberate our proposal from
our ignorance about the Planck-scale physics. To be even more
conservative, we have to check the hierarchy between the
ultraviolet cut-off, \newpage\noindent $\Ld$, of the effective
theory and the inflationary scale. The former can be found from
the non-renormali- zable terms arising in \Eref{action}, whereas
the latter is represented by $\Ve(\sg_*)^{1/4}$ or, less
restrictively, by the corresponding Hubble parameter, $\widehat
H_*=\Ve(\sg_*)^{1/2}/\sqrt{3}\mP$. In particular, the validity of
the effective theory implies \cite{cutoff}
\beq \label{Vl}\mbox{\ftn\sf (a)}\hspace*{0.2cm}
\Ve(\sg_*)^{1/4}\leq\Ld
\hspace*{0.2cm}\mbox{or}\hspace*{0.2cm}\mbox{\ftn\sf
(b)}\hspace*{0.2cm} \widehat H_*\leq\Ld.\eeq
This requirement applies mainly in cases where the involved in
$f(\sg)$ constant $\ck$ takes relatively large values -- as for SM
non-MI \cite{sm1,nmc,shw} --  jeopardizing, thereby, the validity
of the classical approximation, on which the analysis of the
inflationary behavior in this section is based.

\section{Non-Minimal Chaotic Inflation}
\label{ci}

We focus on CI driven primarily by a quadratic potential of the
form
\beq\label{Vci}
V={1\over2}m^2\sg^2+V_{\text{rc}}\hspace*{0.2cm}\mbox{where}\hspace*{0.2cm}V_{\text{rc}}
= \frac{1}{64\pi^2}m^4\ln\frac{m^2}{Q^2}\eeq
are radiative corrections \cite{coleman} to the inflationary
potential. The bulk of our results -- see \Sref{ci2} -- are
independent of the renormalization scale, $Q$, which is set equal
to $\mP$. We below recall (\Sref{ci1}) the results for MCI (with
$f(\sg)=1$) and describe (\Sref{ci2}) our findings for non-MCI,
adopting the following coupling function -- recall that
$\sgb=\sg/\mP$:
\beq\label{fci}
f(\sg)=\lf{1+\ck\sgb}\rg^{-n}\hspace*{0.2cm}\mbox{with}\hspace*{0.2cm}n=\pm1.\eeq

Note, in passing, that results for non-MCI with quartic potential
($V=\ld\sg^4/4!$) are presented in \cref{wmap3, nmchaotic, ld}.
The inflationary scenario based on this potential with $f(\sg)=1$
seems to be excluded \cite{wmap,circ} due to the enhanced
predicted $r$. As we explicitly verified, if we employ the
standard non-minimal coupling function, $f(\sg)=1+\ck\sgb^2$, with
$80\lesssim\ck\lesssim300$ and
$0.2\lesssim\lambda/10^{-4}\lesssim3$ -- c.f. \cref{nmchaotic, ld}
-- we can rescue the model consistently with the constraints of
\Sref{obs} for an indicative $\Trh=10^{10}~\GeV$. In particular
the lower [upper] bound of the allowed regions of $\ck$ and $\ld$
comes from \sEref{subP}{b} [\sEref{Vl}{a} with $\Ld=\mP/\ck$].
Note, however, that the standard non-trivial $f(\sg)$ does not
support reheating along the lines of \Eref{GTrh}.

\subsection{Results for MCI}\label{ci1}

For MCI the slow-roll parameters and the number of $e$-foldings
suffered from $k_*$ can be calculated applying \Eref{sr} and
\Eref{Nhi} -- after removing hats and setting $J=1$ -- with
results
\beq \label{mci1}\epsilon=\eta=2/\sgb^2\hspace*{0.2cm}
\mbox{and}\hspace*{0.2cm}N_*=\lf\sgb_*^2-\sgb_{\rm f}^2\rg/4.\eeq
Using these results, imposing the condition of \Eref{sr} and
employing \Eref{ns} we can derive
\beq \sgb_{\rm f}=\sqrt{2},\hspace*{0.2cm}\sgb_*\simeq2\sqrt{N_*},
\hspace*{0.2cm}\ns\simeq1-{2/ N_*}\hspace*{0.2cm}
\mbox{and}\hspace*{0.2cm} r\simeq8/N_*.\label{mci2}\eeq
Clearly \trns\ values of $\sg$ are required and observationally
favored $\ns$ and $r$ are obtained. More precisely, imposing the
requirements {\ftn \sf (a)} and {\ftn \sf (b)} of \Sref{obs} for
several $\Trh$'s we get numerically the values of $\sg_*$, $m$,
$\ns,~\as$ and $r$ listed in Table~\ref{table} -- c.f.
\cref{circ}. As $\Trh$ decreases, $N_*$ decreases too -- see
\Eref{Ntot} -- and so, $\sg_*$ and $\ns$ slightly decrease whereas
$r$ increases -- see \Eref{mci2}. The resulting $\ns$, $\as$ and
$r$ lie within the range of \Eref{nswmap}. In all cases,
\sEref{subP}{a} is valid whereas the upper bound of
\sEref{subP}{b} is surpassed.

\renewcommand{\arraystretch}{1.2}
\begin{table}[!t]
\begin{center}
\begin{tabular}{|c|cc|ccc|}\hline
$\Trh~(\GeV)$&$\sigma_*/\mP$&$m~(10^{13}~\GeV)$&$\ns$&$\as~(10^{-4})$&$r$\\\hline\hline
$10^{10}$&$15.13$&$1.6$&$0.965$&$6.1$&$0.139$\\
$10^{6}$&$14.73$&$1.69$&$0.963$&$6.7$&$0.147$\\
$10^{5}$&$14.61$&$1.72$&$0.962$&$7$&$0.15$\\
$10^{4}$&$14.5$&$1.74$&$0.962$&$7.2$&$0.152$\\
\hline
\end{tabular}
\end{center}
\vchcaption[]{\sl \small Values of parameters allowed by
Eqs.~(\ref{Prob}), (\ref{Nhi}) and (\ref{Ntot}) for MCI with
several $\Trh$'s.}\label{table}
\end{table}
\renewcommand{\arraystretch}{1.}

\subsection{Results for non-MCI}\label{ci2}

From \eqss{sr}{mci1}{mci2} we can infer that the amplitude of the
inflaton field within non-MCI can become \sub if $J\simeq
1/f(\sg)\gg1$ and $\Ve_{,\sg}/\Ve\simeq V_{,\sg}/V$. These two
objectives can be achieved if we employ $f(\sg)$ given by
\Eref{fci} with $n>0$ and $\ck\gg1$. Another possibility would be
to take $f(\sg)=\exp{\lf-\ck\sgb\rg}$ with $\ck\sim10$. However,
in the latter case the resulting $r$ violates \sEref{nswmap}{c}
and therefore, this option can be declined. Similar problem arises
also if we use $n>1$ -- see \Sref{ci2a}. On the other hand, for
$n=-1$, $\Ve$ in \Eref{VJe} becomes very flat for sufficiently
large $\sgb$'s and so, a new type of non-MCI can takes place.
Decreasing $n$ for $n<0$ we find inflationary solutions, only for
$\ck<0.001$, which break \sEref{subP}{b}. Similar conclusions are
also drawn for the standard non-minimal $f(\sg)$ -- see
\cref{nmchaotic}.

In our numerical code we use as input parameters $m, \sigma_*,
\ck$ and $n$. For every chosen $n$ and $\ck$ we restrict $m$ and
$\sigma_*$ so as the conditions {\ftn\sf (a)} and {\ftn\sf (b)} of
\Sref{obs} -- with $\Trh$ evaluated consistently with \Eref{GTrh}
-- are fulfilled. Our results for $n=+1$ and $n=-1$ are presented
respectively in \Fref{fig1} and \Fref{fig1b}, where we draw the
allowed values of $m$ (solid line) and $\Trh$ (dashed line)
[$\sg_{\rm f}$ (solid line) and $\sg_*$ (dashed line)] versus
$\ck$ for non-MCI (a) [(b)]. For both $n=\pm1$, satisfying
\sEref{subP}{b} gives a lower bound on $\ck$ -- see
\sFref{fig1}{b} and \sFref{fig1b}{b}. On the other hand, the upper
bound on $\ck$ comes from \sEref{nswmap}{c} for $n=+1$ and from
the fact that the enhanced resulting $m$'s destabilize the
inflationary path through the radiative corrections in \Eref{Vci}
for $n=-1$. From our data we also remark that the resulting $m$'s
are almost two orders of magnitude lower [larger] than those
obtained within MCI for $n=+1$ [$n=-1$]. These results depend,
though very weakly, on $\widehat{N}_{\rm tot}$ and therefore, on
the reheating mechanism -- see \Eref{Ntot}. All in all, we obtain
\bea\label{res}
625\lesssim\ck\lesssim2.1\cdot10^7,\hspace*{0.2cm}47\gtrsim
{m\over10^{7}~\TeV}\gtrsim1.6\hspace*{0.2cm}
\mbox{and}\hspace*{0.2cm} 52\gtrsim\widehat
N_*\gtrsim47.9,\hspace*{0.2cm}\mbox{for}\hspace*{0.2cm}n=-1,\\
83\lesssim\ck\lesssim3120,\hspace*{0.2cm}3\lesssim
{m\over10^{12}~\TeV}\lesssim8.6\hspace*{0.2cm}
\mbox{and}\hspace*{0.2cm}58.8\lesssim\widehat
N_*\lesssim59.9,\hspace*{0.2cm}\mbox{for}\hspace*{0.2cm}n=+1.\eea
In both cases, the predicted $\ns$ and $r$ lie within the allowed
ranges of \sEref{nswmap}{a} and \sEref{nswmap}{c} respectively,
whereas $\as$ remains quite small. Our numerical results can be
interpreted through some simple analytical expressions which are
presented in \Sref{ci2a} [\Sref{ci2b}] for $n=+1$ [$n=-1$]. There,
we also comment on the naturalness of our models, following the
arguments of \cref{cutoff, john}.

\subsubsection{Non-MCI with $n=+1$}\label{ci2a}

\begin{figure}[!t]\vspace*{-.21in}
\hspace*{-.19in}
\begin{minipage}{8in}
\epsfig{file=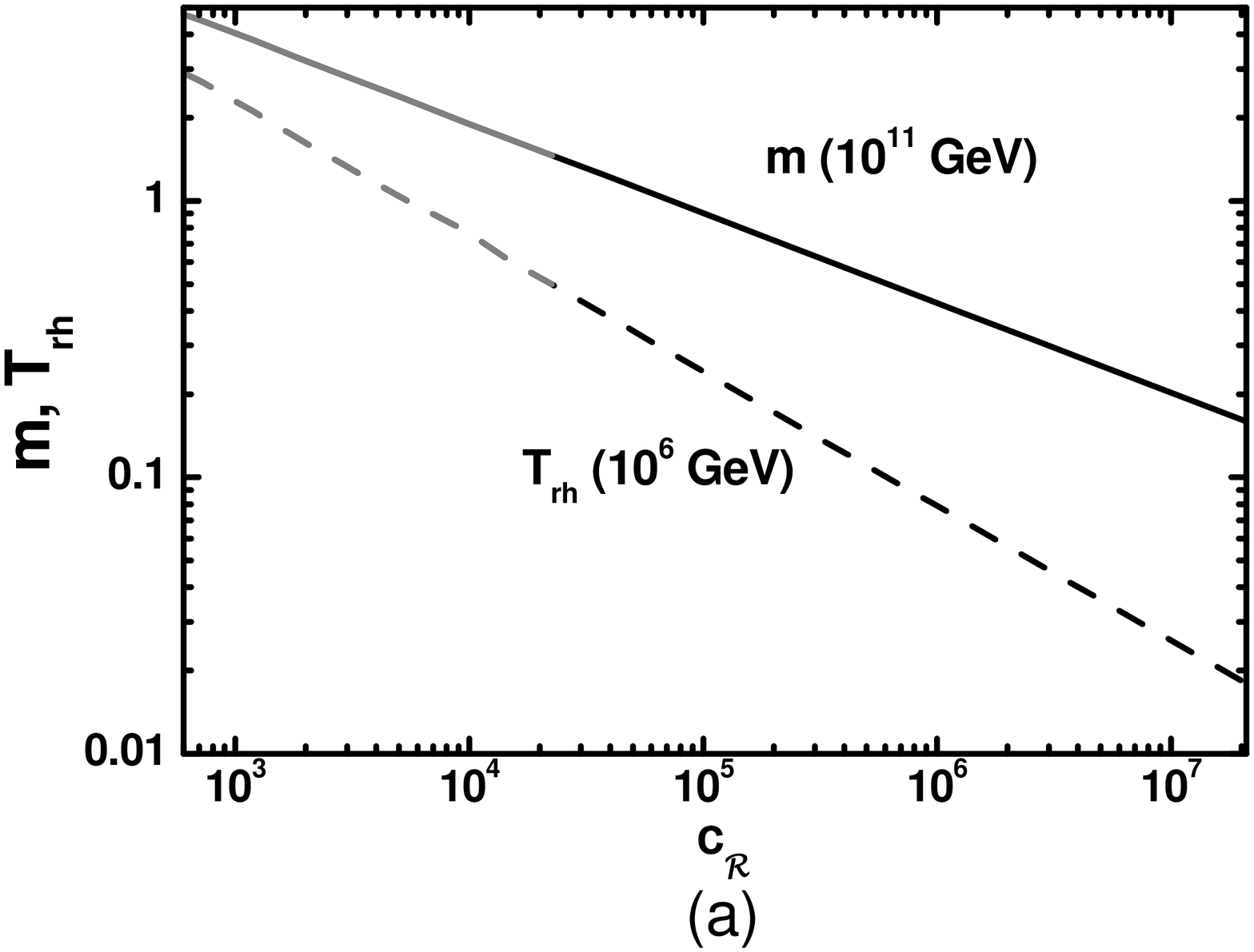,height=3.6in,angle=-90}
\hspace*{-1.2cm}
\epsfig{file=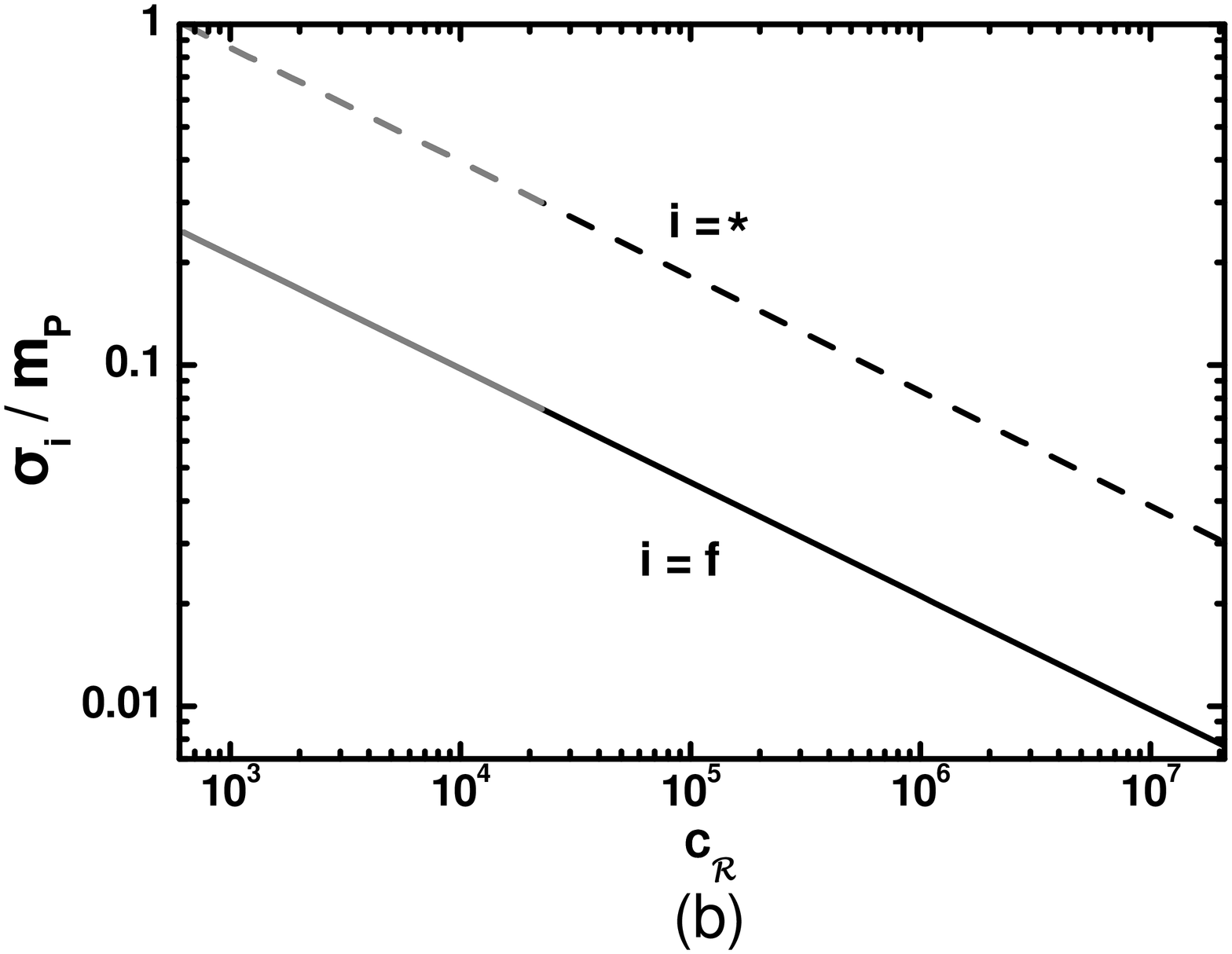,height=3.6in,angle=-90} \hfill
\end{minipage}
\hfill \vchcaption[]{\sl\small  The allowed by Eqs.~(\ref{GTrh}),
(\ref{Prob}), (\ref{Nhi}) and  (\ref{Ntot}) values of $m$ (solid
line) and $\Trh$ (dashed line) [$\sg_{\rm f}$ (solid line) and
$\sg_*$ (dashed line)] versus $\ck$ for non-MCI with $n=-1$ (a)
[(b)]. The gray segments denote values of the various quantities
fulfilling \sEref{Vl}{b} too.}\label{fig1}
\end{figure}

To justify our choice for the negative exponent in \Eref{fci} we
present our formulae below for a general $n>0$. Substituting
\Eref{fci} into \eqs{VJe}{GTrh} and taking into account that
$\ck\gg1$, we obtain
\beq\label{nmci1}
J\simeq\sqrt{\ck^n\sgb^n},\hspace*{0.2cm}\Ve={1\over2}m^2\sg^2\left(1+\ck{\sg\over\mP}\right)^{2n}\simeq
{m^2\ck^{2n}\sg^{2(1+n)}\over2\mP^{2n}}\hspace*{0.2cm}\mbox{and}\hspace*{0.2cm}\Gm{\sg}
\simeq{1\over192\pi}{m_\sg^3\over\mP^2}\eeq
where $m_\sg=m$ and obviously $\sig=0$. Upon use of
\eqss{sr}{Nhi}{nmci1}, the slow roll parameters and
$\widehat{N}_*$ read
\beq\label{nmci2} \hspace*{-0.2cm}\mbox{\ftn\sf
(a)}\hspace*{0.2cm}\widehat\epsilon
\simeq{2(1+n)^2\over\ck^n\sgb^{n+2}},\hspace*{0.2cm}\widehat\eta
\simeq{2(1+n)(1+2n)\over\ck^n\sgb^{n+2}}
=\frac{(1+2n)}{(1+n)}\widehat\epsilon\hspace*{0.2cm}\mbox{and}\hspace*{0.2cm}\mbox{\ftn\sf
(b)}\hspace*{0.2cm} \widehat{N}_{*}\simeq
\frac{\ck^n\left(\sgb^{n+2}_*-\sgb^{n+2}_{\rm
f}\right)}{2(1+n)(2+n)}\cdot\eeq
Imposing the condition of \Eref{sr} and solving then
\sEref{nmci2}{b} w.r.t $\sg_*$ we arrive at
\beq\label{nmci4}\sgb_{\rm
f}\simeq\big(2(1+2n)(1+n)/\ck^n\big)^{1/(n+2)}\hspace*{0.2cm}\mbox{and}\hspace*{0.2cm}
\sgb_*\simeq\lf{2(1+n)(2+n)\widehat{N}_*/\ck^n}\rg^{1/(n+2)}\cdot
\eeq
Inserting the last results into \sEref{nmci2}{a}, we find through
\Eref{ns}
\beq \label{nmci5}\mbox{{\sf\ftn
(a)}}\hspace*{0.2cm}\ns\simeq1-3\widehat\epsilon_*=1-{3(1+n)/
(2+n)\widehat
N_*}\hspace*{0.2cm}\mbox{and}\hspace*{0.2cm}\mbox{{\sf\ftn
(b)}}\hspace*{0.2cm} r\simeq{16(1+n)/(2+n)\widehat N_*}.\eeq
Letting $\ck$ vary within its allowed region for $n=+1$ -- see
\Fref{fig1} -- we find $\ns\simeq(0.952-0.955)$ and
$r\simeq(0.2-0.22)$. Clearly, increasing $n$ leads $r$ above the
range of \sEref{nswmap}{c}. Therefore, we hereafter concentrate on
$n=+1$ which assures an observationally safe and, at the same
time, exciting $r$.

Comparing our findings with those obtained for MCI -- see
Table~\ref{table} -- we notice that the resulting here $\ns$'s are
a little lower, whereas $r$ is significantly elevated and can be
probed in the near future from the measurements of PLANCK
satellite \cite{planck}. Note, in passing, that the so-called Lyth
bound \cite{lyth} on the $\sg$ variation, $\Delta\sg$, gets
modified within non-MI. Namely, combining \eqs{Nhi}{sr} we find
\beq \frac{d\sg}{d\widehat{N}}=\sqrt{r\over8}{\mP\over
J}~~\Rightarrow~~\Delta\sg=\sqrt{r\over8}{\mP\over J}\,\Delta
\widehat{N}~~\Rightarrow~~\Delta\sg\gtrsim\sqrt{2r}\,{\mP\over
J}\simeq\lf2.5-0.083\rg\mP/100,\eeq
taking \cite{lyth} $\Delta \widehat{N}\simeq\Delta N=4$ and
assuming negligible variation of $f(\sg)$ from its value at
$\sg=\sg_*$. Therefore, large $r$'s do not correlate necessarily
with \trns\ $\Delta\sg$'s within non-MI. On the other hand, $\se$
as evaluated from \Eref{VJe}, $\se\simeq \sqrt{\ck\sg^3/\mP}$,
remains \trns.

%%%%%%%%%%%%%%%%%%%%%%%%%%%%%%%%%%%%%%%%%%%%%%%%%%%%%%%%%%%%%%%%%%%%%

%%%%%%%%%%%%%%%%%%%%%%%%%%%%%%%%%%%%%%%%%%

The resulting $\Ve$ in \Eref{nmci1} is non-renormalizable and
suggests that the theory breaks down for energies of the order
$\Ld=\mP/\ck$. Checking the consistency with \sEref{Vl}{a} we find
numerically:
\beq\label{ress2}0.03\lesssim\widehat H_*/\Ld\lesssim1
\hspace*{0.2cm}\mbox{for}\hspace*{0.2cm}625\lesssim\ck\lesssim2.26\cdot10^4
\hspace*{0.2cm}\mbox{and}\hspace*{0.2cm} 1\gtrsim\sgb_*\gtrsim
0.3,\eeq
where the corresponding ranges of values are depicted by the gray
segments of the lines in \Fref{fig1}. The range in \Eref{ress2}
turns out to be a little more comfortable than the one we get
within SM non-MI -- c.f. \cref{cutoff}. However, \sEref{Vl}{a} is
violated, since $\Ve(\sg_*)^{1/4}/\Ld\gtrsim5.8$.

On the other hand, non-renormalizable terms in the action of
\Eref{action1} and (\ref{action}) indicate that $\Ld=\mP$. In
fact, such terms arise from the first term in \Eref{action1} and
the second one in \Eref{VJe}. The form of these terms is generated
expanding the relevant coefficients in series around $\sg=\sg_*$
with the following result -- an expansion in the small field
limit, $\ck\sgb\ll1$, fails to reproduce the exact results:
\vspace{-0.0cm} \numparts\baq \label{ress1}
\mP^2f\rcc&\ni&{\mP\over\ck\sgb_*}\lf 1-3
{\sgb\over\sgb_*}+10\lf{\sgb\over\sgb_*}\rg^2+\cdots\rg\partial^{\bar\mu}\partial_{\bar\mu}
h^{\mu\nu} \\\mbox{and}\hspace*{0.2cm} \mP^2{f_{,\sg}^2\over f^2}
\geu^{\mu\nu}\partial_\mu\sg\partial_\nu\sg&\ni&{1\over\sgb_*^2}\lf1-
8{\sgb\over\sgb_*}+45\lf{\sgb\over\sgb_*}\rg^2+\cdots\rg
\geu^{\mu\nu}\partial_\mu\sg\partial_\nu\sg,\label{ress3}\eaq\endnumparts
\hspace{-.14cm}
where $h^{\mu\nu}$ denotes the graviton field involved in the
expansion \cite{cutoff, john} of the metric
$g_{\mu\nu}\simeq\eta_{\mu\nu}+h_{\mu\nu}/\mP$ around the
Minkowski space with metric $\eta_{\mu\nu}$ and \rcc\ is
approximated linearly. Given these ambiguities, we do not consider
\Eref{ress2} as absolute constraint.

\subsubsection{Non-MCI with $n=-1$}\label{ci2b}

%%%%%%%%%%%%%%%%%%%%%%%%%%%%%%%%%%%%%%%%%%%%%%%%%%%%%%%%%%%%%%%%%%%%%
\begin{figure}[!t]\vspace*{-.21in}
\hspace*{-.19in}
\begin{minipage}{8in}
\epsfig{file=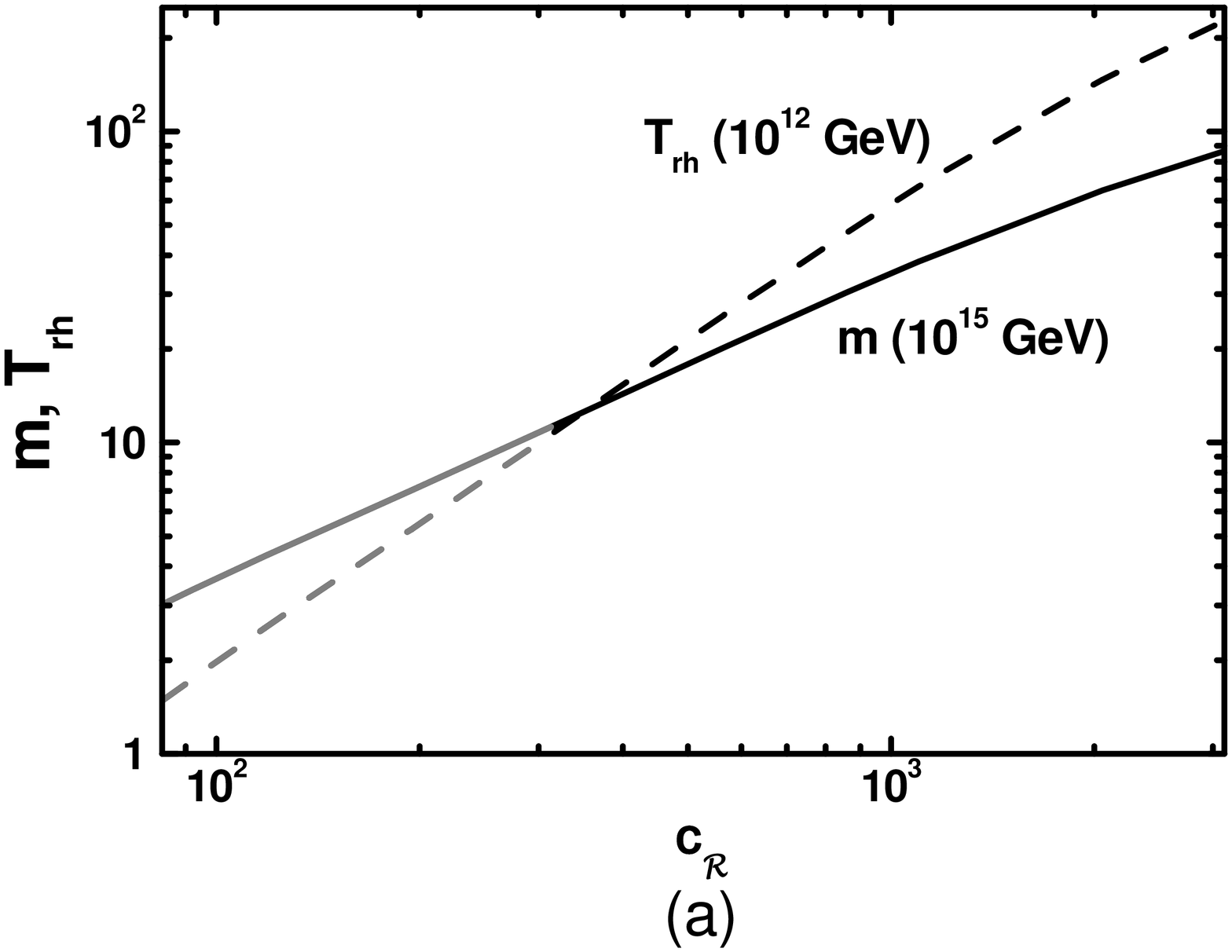,height=3.6in,angle=-90}
\hspace*{-1.2cm}
\epsfig{file=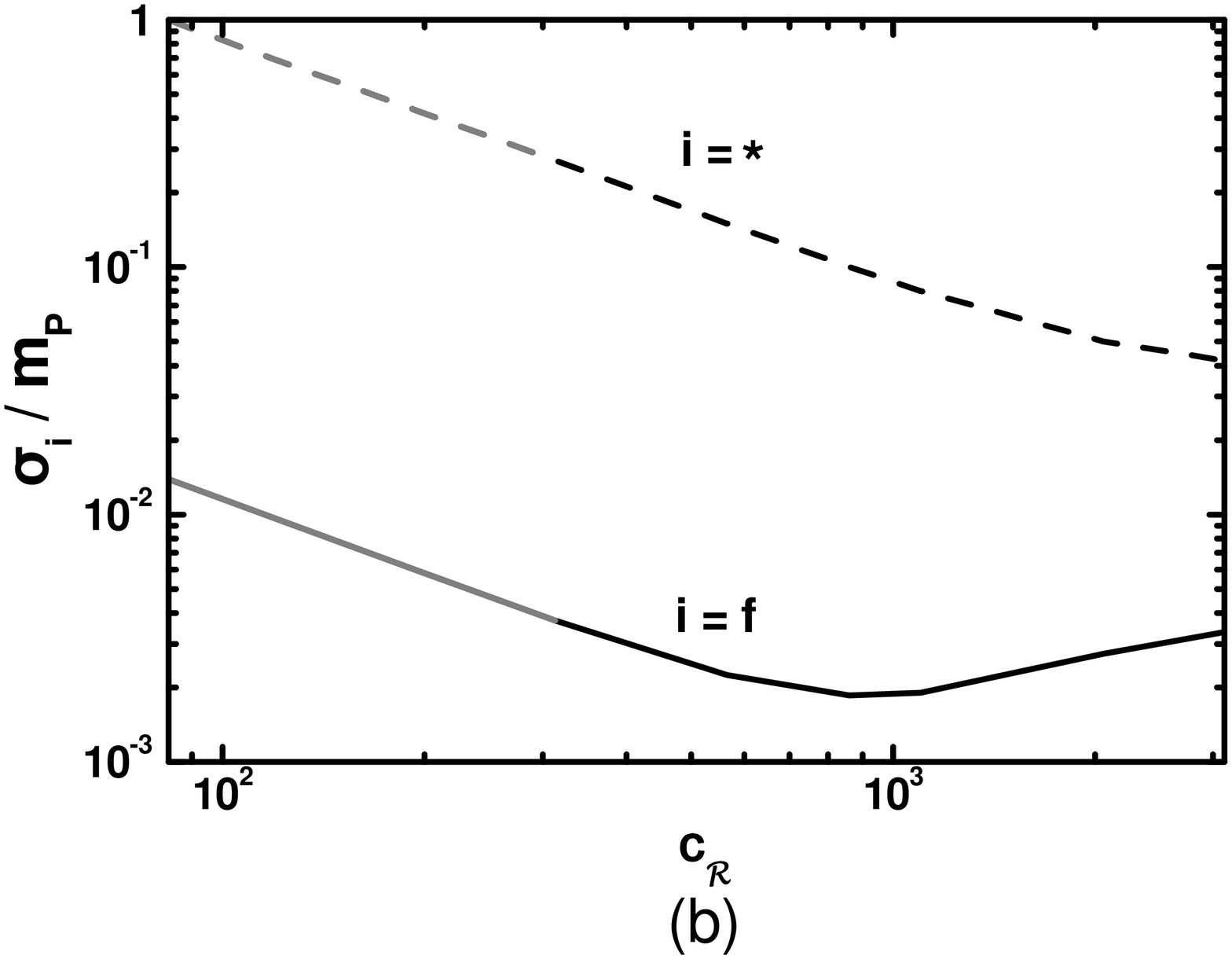,height=3.6in,angle=-90} \hfill
\end{minipage}
\hfill \vchcaption[]{\sl\small  The same as in \Fref{fig1} but for
$n=-1$. The gray segments here denote values of the various
quantities fulfilling \sEref{Vl}{a} too.}\label{fig1b}
\end{figure}

%%%%%%%%%%%%%%%%%%%%%%%%%%%%%%%%%%%%%%%%%%

A completely different situation from that studied in \Sref{ci2a}
emerges for $n=-1$ in \Eref{fci}. Indeed, substituting \Eref{fci}
into \eqs{VJe}{GTrh} and taking into account that $\ck\gg1$, we
obtain
\beq\label{nmci1b} J\simeq\sqrt{3/2}\
\sgb^{-1},\hspace*{0.2cm}\Ve\simeq
{m^2\mP^2\over2\ck^2}\hspace*{0.2cm}\mbox{and}\hspace*{0.2cm}\Gm{\sg}
\simeq{1\over192\pi}{m_\sg^3\over\mP^2}\eeq
where $m_\sg=m$ and obviously $\sig=0$. We observe that $\Ve$
exhibits a flat plateau as we obtain for the quatric potential
with the standard non-minimal $f(\sg)$ -- c.f.
\cref{sm1,nmc,shw,ld}. Employing \eqss{sr}{Nhi}{nmci1b}, the slow
roll parameters and $\widehat{N}_*$ read
\beq\label{nmci2b} \mbox{\ftn\sf
(a)}\hspace*{0.2cm}\widehat\epsilon
\simeq{4\over3\ck^2\sgb^2},\hspace*{0.2cm}\widehat\eta
\simeq-{4\over3\ck\sgb}
=-\widehat\epsilon\ck\sgb\gg-\widehat\epsilon\hspace*{0.2cm}\mbox{and}\hspace*{0.2cm}\mbox{\ftn\sf
(b)}\hspace*{0.2cm} \widehat{N}_{*}\simeq
\frac{3\ck}{4}\left(\sgb_*-\sgb_{\rm f}\right).\eeq
As opposed to our findings in \eqs{mci1}{nmci2}, notice that
$\eta<0$ here. Imposing the condition of \Eref{sr} and solving
then \sEref{nmci2b}{b} w.r.t $\sg_*$ we arrive at
\beq\label{nmci4b}\sgb_{\rm
f}\simeq2/\sqrt{3}\ck\hspace*{0.2cm}\mbox{and}\hspace*{0.2cm}
\sgb_*\simeq4\,\widehat{N}_*/3\ck\cdot \eeq
Inserting the last results into \sEref{nmci2b}{a}, we find through
\Eref{ns}
\beq \label{nmci5b}\mbox{{\sf\ftn
(a)}}\hspace*{0.2cm}\ns\simeq1+2\widehat\eta_*=1-{2/\widehat
N_*}\simeq(0.967-0.97)\hspace*{0.2cm}\mbox{and}\hspace*{0.2cm}\mbox{{\sf\ftn
(b)}}\hspace*{0.2cm} r\simeq{12/\widehat
N^2_*}\simeq(0.002-0.003),\eeq
where the ranges above are derived numerically letting $\ck$ vary
within its allowed region -- see \Fref{fig1b}. Notice that the
resulting $\ns$'s and $r$'s are identical to those derived in
\cref{sm1,shw}. Comparing them with those listed in
Table~\ref{table} or given in the paragraph below \Eref{nmci5} we
remark that $r$ is significantly reduced, whereas $\ns$ is close
to the value obtained in MCI and a bit larger than the one
extracted for non-MCI with $n=+1$.

As for the latter case, non-renormalizable terms in the action of
\Eref{action1} indicate an effective cutoff $\Ld=\mP/\ck$, since
\beq \label{ress1b} \mP^2f\rcc\ni{\ck\over\mP}\ \sg\
\lf\partial^{\mu}h_{\mu\bar\mu}\partial^{\nu}
h_{\nu}^{\bar\mu}+\partial^{\bar\mu}h_{\mu\nu}\partial^{\mu}
h_{\bar\mu\nu}+\partial_{\mu}h\partial^{\mu}h \rg\eeq
with $h=h_\mu^\mu=h_{\mu\mu}$. On the other hand, the second term
in \Eref{VJe} gives exactly the same result as in \Eref{ress3}
since $f_{,\sg}/f$ is identical for both $n=\pm1$ in \Eref{fci}.
Checking the consistency with \sEref{Vl}{a} we find numerically
\beq\label{ress2b}0.3\lesssim\widehat V(\sg_*)^{1/4}/\Ld\lesssim1
\hspace*{0.2cm}\mbox{for}\hspace*{0.2cm}83\lesssim\ck\lesssim313
\hspace*{0.2cm}\mbox{and}\hspace*{0.2cm} 1\gtrsim\sgb_*\gtrsim
0.27,\eeq
where the corresponding ranges of values are depicted by the gray
segments of the lines in \Fref{fig1b}. On the other hand,
\sEref{Vl}{b} is satisfied in the whole parameter space of these
figures. Consequently, non-MCI with $n=-1$ can be characterized as
more natural than the one with $n=+1$.

\paragraph{} Concluding this section, let us emphasize that, in contrast to the models
suggested in \cref{circ}, non-MCI is not of hilltop type and so,
complications related to the initial conditions are avoided.
Indeed, for non-MCI with $n=+1$, $\Ve$ remains concave upwards,
whereas with $n=-1$, $\Ve$ develops a plateau without
distinguished maximum. As we explicitly checked, possible
inclusion of extra radiative corrections in \Eref{Vci} due to a
coupling of $\sg$ to fermions -- considered in \cref{circ} -- do
not affect our proposal for values of the relevant Yukawa coupling
constant, $h$, lower than about $10^{-3}$. For such $h$'s, the
decay width of the inflaton due to this channel dominates over the
one given by \Eref{GTrh}.

\section{Non-Minimal Hybrid Inflation}\label{hi}

Hybrid inflation can be realized in the presence of two real
scalar fields, $\sg$ and $\phi$, involved in the following
potential \cite{hybrid}
\begin{equation}\label{Vhi1}
V \left(\phi,\sigma\right) = \kappa^2 \left(M^2 -
\frac{\phi^2}{4}\right)^2+\frac{m^2\sigma^2}{2}+\frac{\lambda^2
\phi^2 \sigma^2}{4},
\end{equation}
where $M$, $m$ are mass parameters and $\kappa$, $\lambda$ are
dimensionless coupling constants. The global minima of $V$ lie at
$\left(\sig,\langle \phi \rangle\right)= \left(0,\pm 2M\right)$.
Therefore, $V$ leads to a spontaneous symmetry breaking of a
global or local symmetry depending on the nature of the waterfall
field $\phi$. In the latter case, topological defects may be also
produced via the Kibble mechanism \cite{kibble}. Trying to keep
our approach as simple as possible we below assume that this is
not the case.

In addition, $V$ in \Eref{Vhi1} gives rise to HI. This is because
$V$ possesses an almost $\sg$-flat direction  at $\phi = 0$ with
constant potential density equal to
$V_0=V\lf\phi=0,\sg\rg=\kappa^2M^4$, for $m=0$. The effective mass
squared of the field $\phi$ along this direction is
\beq \label{sc} m^2_{\phi} = -\kappa^2 M^2 +
\lambda^2\sg^2/2>0~~\Leftrightarrow~~\sg
> \sg_c = \sqrt{2}\kappa M/\lambda\,.\eeq
Thus, for $\sg> \sg_c$ the $\phi = 0$ direction represents a
valley of minima which can serve as inflationary trajectory. On
this path the potential of HI takes the form
\beq V=V_0+{1\over2}m^2\sg^2+V_{\rm rc}\label{Vhi}\eeq
where $V_{\rm rc}$ is the one-loop correction (to the tree-level
potential) which can be written as \cite{coleman, hirc}
\numparts\baq \label{Vrc} V_{\text{rc}} &=&
\frac{1}{64\pi^2}\left(m^4\ln\frac{m^2}{Q^2}+
\kappa^2V_0\left(x-1\right)^2\ln\frac{\kappa^2M^2}{Q^2}\left(x-1\right)\right)
~~\mbox{with}~~x=\left({\sg\over\sgc}\right)^2\\\label{Vrc1}
&\simeq&
\frac{\kappa^2V_0}{64\pi^2}\left(x^2\ln\frac{\kappa^2M^2}{Q^2}+{3\over2}\right)~~\mbox{for}~~x\gg1.
\eaq\endnumparts \hspace{-.14cm}
Here, $Q$ is a renormalization scale which can be conveniently
chosen \cite{hirc} equal to $\sgc$ which practically coincides
with the value of $\sg$ at the end of HI, $\sg_{\rm f}$, for both
MHI and non-MHI.

We below review (\Sref{res1}) the results for MHI (with
$f(\sg)=1$) and describe (\Sref{res2}) our findings for non-MHI,
seeking the following non-minimal coupling function for the
inflaton -- for earlier attempts on non-MHI, see \cref{nmhybrid}:
\beq \label{fhi} f(\sg)=1-\ck\sgb/\lf1+\sgb\rg^2\eeq
where we use, as usually, the shorthand $\sgb=\sg/\mP$. As regards
the waterfall field we can assume that it is either minimally
coupled to gravity or its coupling function is $f(\phi)$ since
$f(0)=1$ and $f(2M)\simeq1$ for $\ck\ll1$ and $M\leq\mP$.

\subsection{Results for MHI}\label{res1}

We can get an impression of the expected results for MHI, if we
calculate the, involved in the inflationary dynamics, derivatives
of $V$ in \Eref{Vhi}. Namely we have
\beq\label{Vs} V_{,\sg}=m^2\sg+{x\kappa^2V_0\over 32\pi^2\sg}\lf
x-1\rg\lf2\ln{\kappa^2M^2\over Q^2}\lf x-1\rg+1\rg.\eeq
We observe that there are two contributions in $V_{,\sg}$. The
first one arises from the tree-level potential whereas the second
one comes from the radiative corrections in \Eref{Vrc}. When the
first contribution dominates over the second one, we obtain the
well-known tree-level \cite{hybrid} results, $N_{\rm tr*}$ and
$\eta_{\rm tr}$, for $N_*$ and $\eta$ respectively -- note that we
identify $\sg_{\rm f}$ with $\sgc$:
\beq\label{tree} N_{\rm tr*}={1\over \eta_{\rm
tr}}\ln{\sg_*\over\sg_{\rm
c}}\hspace*{0.2cm}\mbox{with}\hspace*{0.2cm}\eta_{\rm
tr}=\mP^2{m^2\over V_0}\gg\epsilon\,.\eeq
In this regime, the resulting $\ns$ clearly -- see \Eref{ns} --
exceeds slightly unity in contrast to the observationally favored
results of \sEref{nswmap}{a}. Moreover, as we find numerically,
the lower $\kappa$ and/or $m$ we use, the closer $\sigma_*$ is set
to $\sigma_{\rm c}$. This is the first kind of tuning occurred
within MHI.

Nonetheless, taking into account that the logarithm in \Eref{Vs}
turns out to be negative, we can show that, for every $m$, there
is $\kappa$ such that $V$ develops a maximum at
$\sigma=\sigma_{\rm max}$, which can be estimated by numerically
solving the condition $V_{,\sg}(\sigma_{\rm max})=0$. At
$\sigma=\sigma_{\rm max}$, $V_{,\sg\sg}$ given by
\beq\label{Vss}
V_{,\sg\sg}={V_{,\sg}\over\sg}+{x^2\kappa^2V_0\over
16\pi^2\sg^2}\lf2\ln{\kappa^2M^2\over Q^2}\lf x-1\rg+3\rg, \eeq
becomes negative and so, $\eta$ and $\ns$ start decreasing for
$\sigma_*$ close $\sigma_{\rm max}$ -- see \eqs{sr}{ns}. As for
any model of hilltop inflation, the lower $n_{\rm s}$ we obtain,
the closer $\sigma_*$ is located to $\sigma_{\rm max}$. This is a
second kind of tuning which remains even for non-MHI -- see
\Sref{res2}. To quantify somehow the amount of the tunings
encountered in the considered model, we define the quantities:
\beq \mbox{\sf\ftn (a)}\hspace*{0.2cm} \Dex=\frac{\sigma_{\rm max}
- \sigma_*}{\sigma_{\rm
max}}\hspace*{0.2cm}\mbox{and}\hspace*{0.2cm}\mbox{\sf\ftn
(b)}\hspace*{0.2cm}\Dcex=\frac{\sigma_*-\sigma_{\rm
c}}{\sigma_{\rm c}}\cdot\label{dms}\eeq

%%%%%%%%%%%%%%%%%%%%%%%%%%%%%%%%%%%%%%%%%%%%%%%%%%%%%%%%%%%%%%%%%%%%%
\begin{figure}[!t]\vspace*{-.16in}
\hspace*{-.23in}
\begin{minipage}{8in}
\epsfig{file=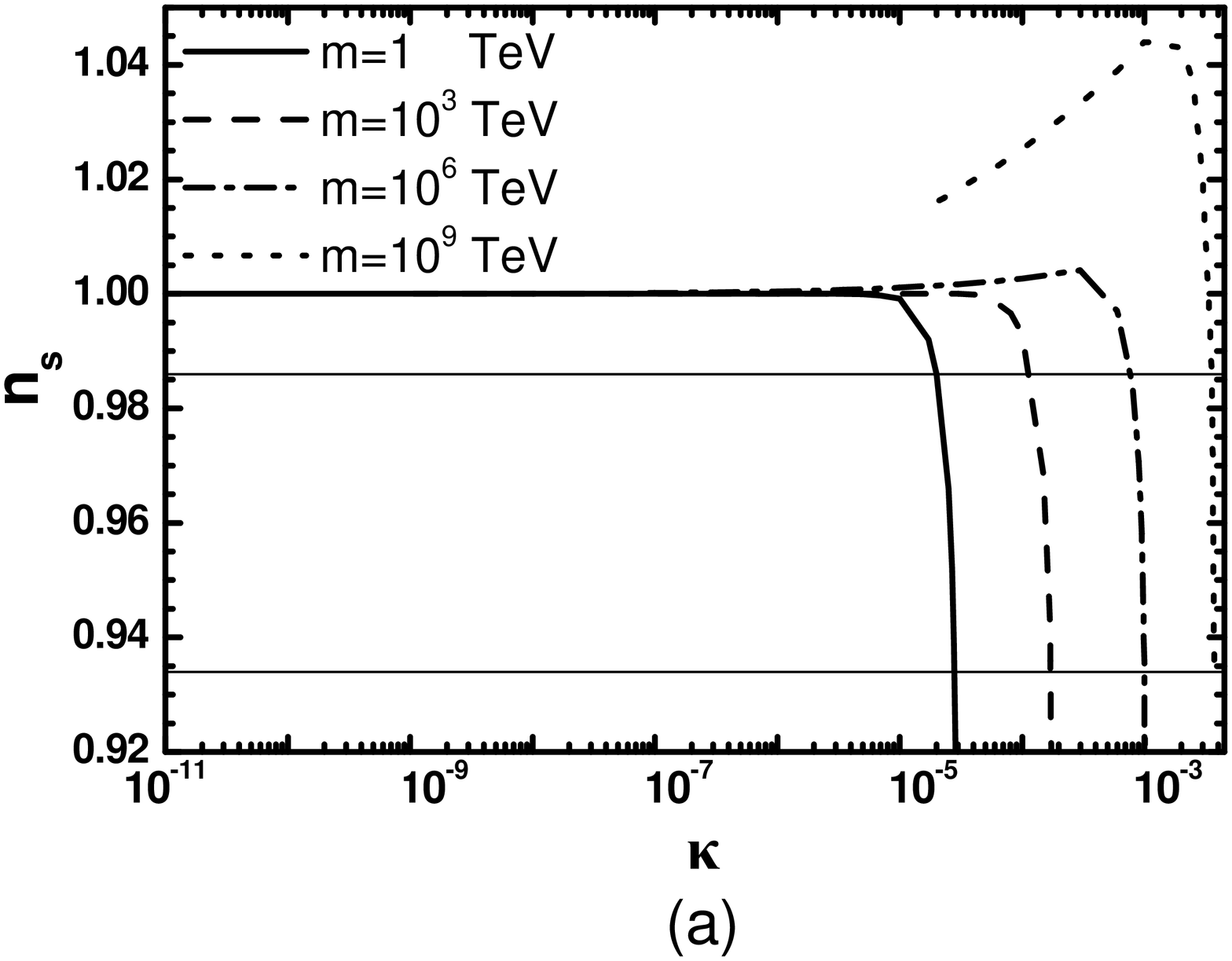,height=3.6in,angle=-90}
\hspace*{-1.2cm}
\epsfig{file=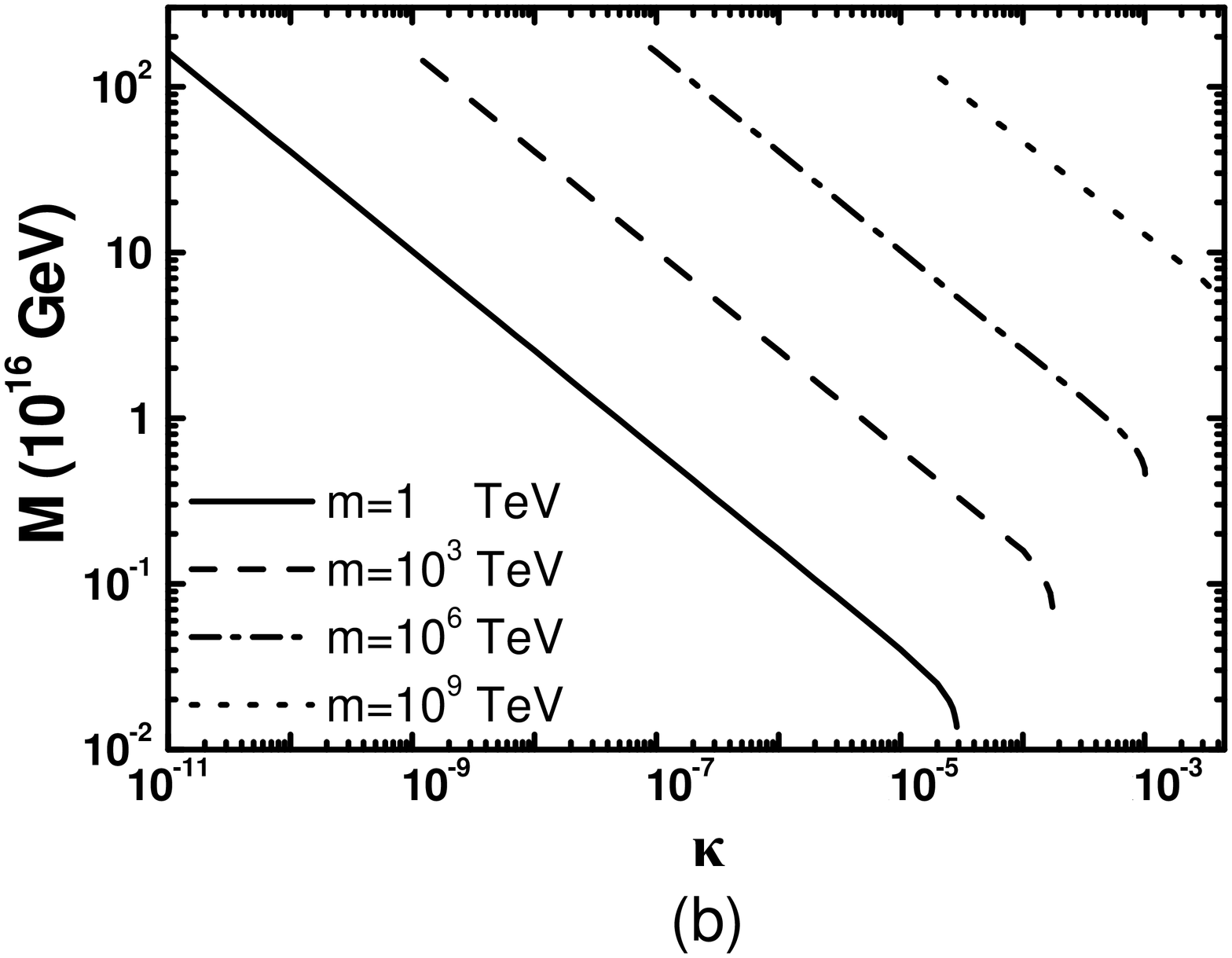,height=3.6in,angle=-90} \hfill
\end{minipage}\hfill\begin{center}\renewcommand{\arraystretch}{1.1}
\begin{tabular}{|c|cc|cc|}\hline
$m~(\TeV)$&$\kappa~(10^{-3})$&$M~(10^{16}~\GeV)$&$\Dcex$&$\Dex$\\\hline\hline
$1$&$0.02 - 0.028$&$0.025 - 0.015$&$0.00016-0.00029$&$0.0004 - 0.00007 $\\
$10^{3}$&$0.12 - 0.17$&$0.13 - 0.083$&$0.0053 - 0.01$&$0.013 - 0.0023$\\
$10^{6}$&$0.8 - 1$&$0.69 - 0.49$&$0.18 - 0.31$&$0.19 - 0.054$\\
$10^{9}$&$3.55 - 3.72$&$5.98 - 5.5$&$4.4 - 5.7$&$0.38 - 0.24$\\
\hline
\end{tabular}
\end{center}\vspace*{-.08in}
\hfill \vchcaption[]{\sl\small  The allowed by
\eqss{Prob}{Nhi}{Ntot} values of $\ns$ {\ssz\sf (a)} and $M$
{\ssz\sf (b)} versus $\kappa$ for MHI with $\Trh=10^{10}~\GeV$,
$\kappa=\lambda$ and several $m$'s indicated in the graphs. Shown
are also in the table the allowed ranges of the various parameters
for $\ns$ in the range of \sEref{nswmap}{\ssz a} limited by thin
lines {\ssz\sf (a)}.}\label{fig2}
\end{figure}\renewcommand{\arraystretch}{1.}

%%%%%%%%%%%%%%%%%%%%%%%%%%%%%%%%%%%%%%%%%

The above rough estimations can be verified by our numerical
computations. In our code, we use as input parameters $\kappa$,
$\ld$, $m$, $M$,  $\sigma_*$ and $\Trh$. In our analysis for MHI,
we fix $\Trh=10^{10}~\GeV$ and $\kp=\ld$ -- possible variation of
these two choices do not modify our conclusions in any essential
way. For any chosen $\kp$ and $m$ we then restrict $M$ and
$\sigma_*$ so as the restrictions {\ftn\sf (a)}, {\ftn\sf (b)} and
{\ftn\sf (d)} of \Sref{obs} and \Eref{sc} are fulfilled. Using
Eq.~(\ref{ns}) we can extract $\ns,~\as$ and $r$. Our results are
presented in \sFref{fig2}{a} [\sFref{fig2}{b}] where we design the
allowed values of $\ns$ [$M$] versus $\kappa$ for $m=1~\TeV$
(solid line) or $m=10^3~\TeV$ (dashed line) or $m=10^6~\TeV$
(dot-dashed line) or $m=10^9~\TeV$ (dotted line). The region of
\sEref{nswmap}{a} is also limited by thin lines. The various lines
terminate at low $\kappa$'s due to the saturation of
\sEref{subP}{b} and at large $\kappa$'s since the imposed
conditions can not be fulfilled.

Clearly, the almost horizontal part of the various lines, which
exceeds the observational limits of \sEref{nswmap}{a}, in the
$\kappa-\ns$ plane corresponds to the dominance of the tree-level
potential. However, for any $m$ and relatively large $\kappa$'s we
can obtain acceptable $\ns$'s even without inclusion of extra
radiative corrections due to a possible coupling of the inflaton
to fermions -- c.f. \cref{hirc}. On the other hand, it is worth
emphasizing that the allowed range of $\kappa$'s for each $m$ is
severely tuned. Indeed, confining $\ns$ within the range of
\sEref{nswmap}{a} we find the ranges of the parameters listed in
the table of \Fref{fig2}. From the outputs there, we also remark
that $\kp$'s, $M$'s, $\Dcex$'s and  $\Dex$'s increase with $m$.
Therefore, the natural realization of MHI requires large $m$'s. In
this case too, $M$ turns out to be well above its value within the
SUSY version HI -- c.f. \cref{susyhybrid, gpp}. Needless to say,
finally, that the resulting $\as$'s and $r$'s turn out to be
vanishingly small and so, uninteresting. In conclusion, MHI (with
the minimal possible radiative corrections) is rather disfavored
by the current observational data.

\subsection{Results for non-MHI} \label{res2}

From the analysis of MHI we can deduce that reduction of $\ns$ for
a wider range of $\kp$'s can be achieved if the slope of $V$
becomes steeper. This objective can be achieved if we employ
$f(\sg)$ given by \Eref{fhi} with $\ck\ll1$. Another possibility
would be $f(\sg)=\exp{\lf-\ck\sgb\rg}$ or that of \Eref{fci} with
$n>0$ and $\ck\sim0.1$. However, in these cases the resulting
$\sg_*$ violates the bound of \sEref{subP}{b} and therefore, these
options are not adoptable. Moreover, imposing on non-MHI with the
standard non-minimal $f(\sg)$ the constraints {\sf\ftn (a)},
{\sf\ftn (b)} and {\sf\ftn (d)} of \Sref{obs} and \Eref{sc}, we
are obliged to use a tiny $\ck\sim-10^{-3}$, which has no sizable
impact on reducing $\ns$. Consequently, this last choice can not
become observationally viable, too.

Differentiating \Eref{fhi} w.r.t $\sg$, substituting into
\eqs{VJe}{GTrh} and taking into account that $\ck\ll1$, we obtain
\beq\label{nmhi1}
f_{,\sg}=\frac{\ck\lf-1+\sgb\rg}{\mP\lf1+\sgb\rg^3},\hspace*{0.2cm}
f_{,\sg\sg}=\frac{2\ck\lf2-\sgb\rg}{\mP^2\lf1+\sgb\rg^4},\hspace*{0.2cm}\Ve\simeq
V_0,\hspace*{0.2cm}J\simeq1\hspace*{0.2cm}\mbox{and}\hspace*{0.2cm}
\Gm{\sg}= {\ck^2\over 128\pi}{m_\sg^3\over\mP^2}\eeq
where $m_\sg=\sqrt{2\lambda^2M^2+m^2}$. Despite the fact that
$\Ve$ given by \eqs{VJe}{Vhi} is practically equal to $V_0$ --
since $f(\sgb)\simeq1$ for $\sgb\ll1$ --, its inclination is
mostly dominated by the term $-2V_0f_{,\sg}$ of $V_{,\sg}$.
Indeed, upon use of \eqss{sr}{Vrc1}{nmhi1} we find
\beq\label{nmhi2}
\widehat\epsilon={\mP^2\over2}\left(-2f_{,\sg}+{m^2\sg\over
V_0}+{\kappa^2x^2\over16\pi^2\sg}\ln{\kp^2M^2\over
Q^2}\right)^2\cdot\eeq
In a sizable portion of the parameter space, the first
contribution to $\widehat\epsilon$ in \Eref{nmhi2} overshadows the
others two. As a consequence, $\Ve$ develops a maximum at
$\sgb=\sgb_{\rm max}$ for $f_{,\sg}\lf\sgb_{\rm
max}\rg=0\Leftrightarrow\sgb_{\rm max}\simeq1$ with
$\Ve_{,\sg\sg}\lf\sg_{\rm max}\rg<0$. In fact, inserting
\eqs{VJe}{Vhi} into \Eref{sr} we end up with
\beq\label{nmhi3} \widehat\eta=\mP^2\left(-2f_{,\sg\sg}+{m^2\over
V_0}+{3\kappa^2x^2\over16\pi^2\sg^2}\ln{\kp^2M^2\over
Q^2}\right),\eeq
which is negative for dominant $f_{,\sg\sg}$ with $\sgb<2$.
Combining Eqs.~(\ref{nmhi3}) with (\ref{ns}{\ftn\sf a}) we can
easily infer that $\ck>0$ for $\sgb<\sgb_{\rm max}$ strengthens
significantly the reduction of $n_{\rm s}$. Neglecting the two
last terms in the right-hand side of \Eref{nmhi2}, we can estimate
$\widehat{N}_*$ via \Eref{Nhi} with result
\beq \label{Nnmhi}\widehat
N_*\simeq{1\over2\mP^2}\int_{\sg_*}^{\sgc}{d\sg\over
f_{,\sg}}={1\over6\ck}\left(\lf21+6\sgb_*+
\sgb_*^2\rg\sgb_*-\lf21+6\sgb_{\rm c}+\sgb_{\rm c}^2\rg\sgb_{\rm
c}+24\ln{1- \sgb_*\over1-\sgb_{\rm c}}\right)\cdot\eeq
As we verify numerically, the formula above gives accurate results
for $m\leq10^6~\TeV$ and sufficiently low $\kp$'s. However, since
$\sg_*$ depends on $\widehat{N}_*$ in a rather complicate way, it
is not doable to find an analytical result for $\ns$ as a function
of $\widehat{N}_*$ -- c.f. \Eref{nmci4} and \Eref{nmci4}.
Therefore, our last resort is the numerical computation, whose the
results are presented in the following.

%%%%%%%%%%%%%%%%%%%%%%%%%%%%%%%%%%%%%%%%%%%%%%%%%%%%%%%%%%%%%%%%%%%%%
\begin{figure}[!t]\vspace*{-.16in}
\hspace*{-.23in}
\begin{minipage}{8in}
\epsfig{file=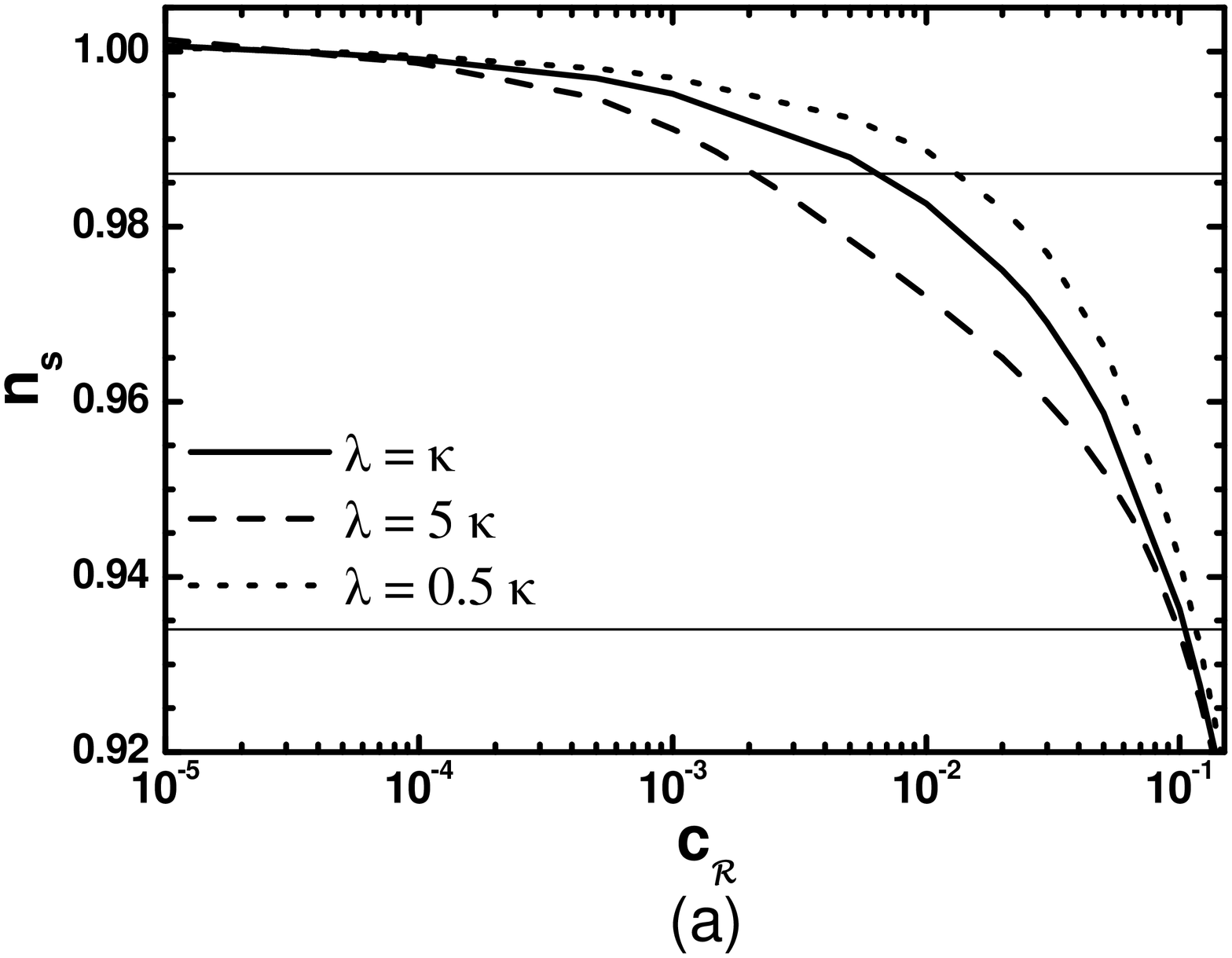,height=3.6in,angle=-90}
\hspace*{-1.2cm}
\epsfig{file=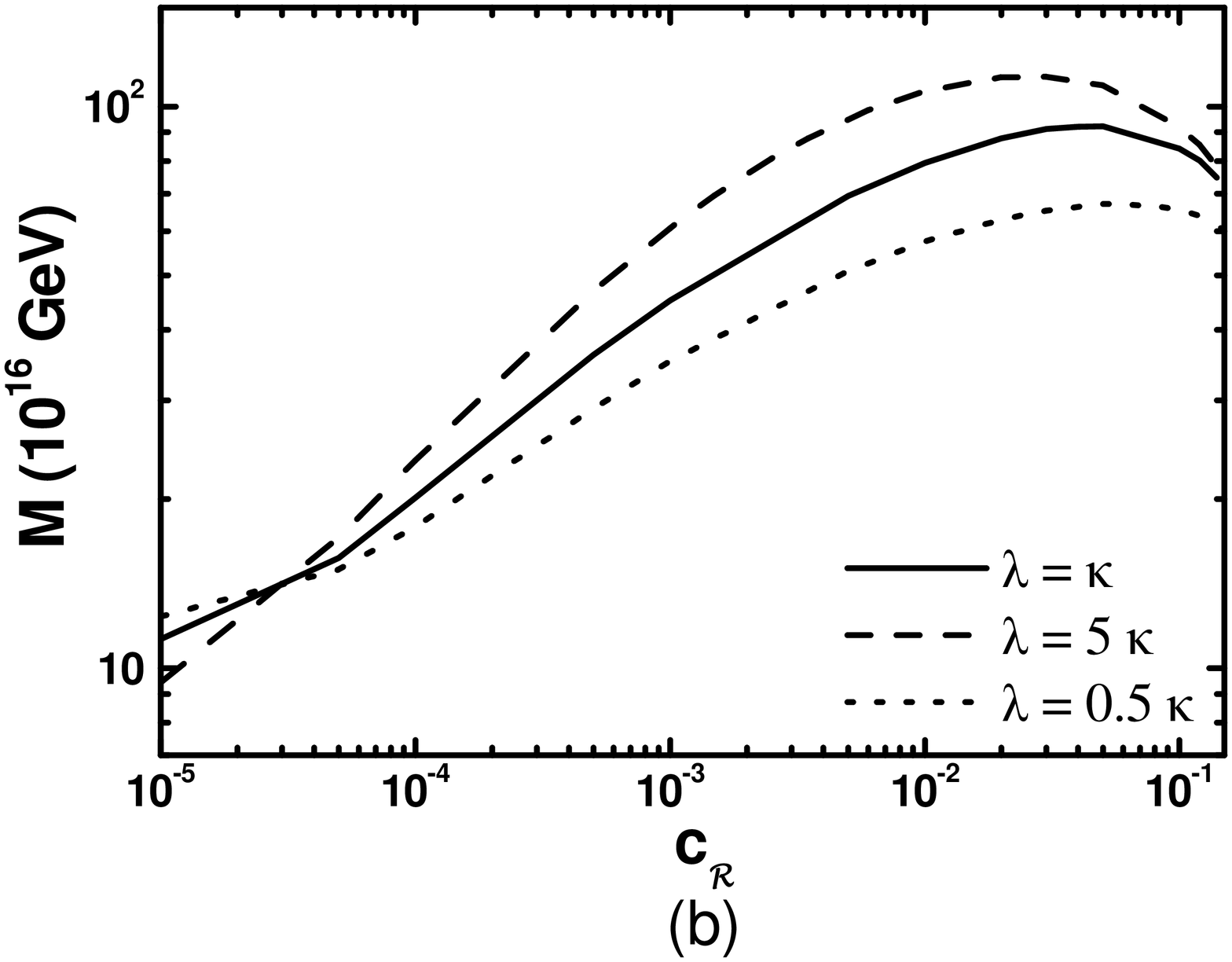,height=3.6in,angle=-90} \hfill
\end{minipage}
\hfill \vchcaption[]{\sl\small The allowed by Eqs.~(\ref{Prob}),
(\ref{Nhi}) and (\ref{Ntot}) -- with $\Trh$ given by \Eref{GTrh}
-- values of $\ns$ {\ssz\sf (a)} and $M$ {\ssz\sf (b)} versus
$\ck$ for non-MHI with $m\leq10^6~\TeV$, $\kappa=10^{-5}$ and
several $\lambda$'s indicated in the graphs. The region of
\sEref{nswmap}{\ssz a} is also limited by thin lines.}\label{fig5}
\end{figure}

%%%%%%%%%%%%%%%%%%%%%%%%%%%%%%%%%%%%%%%%%

In our code, we use as input parameters $\kappa$, $\ld$, $m$, $M$,
$\sigma_*$ and $\ck$. Note that $\Trh$ is calculated via
\Eref{GTrh}. For every chosen $\kappa$, $\ld$, $m$ and $\ck$, we
can restrict $M$ and $\sigma_*$ so as the conditions {\sf\ftn
(a)}, {\sf\ftn (b)} and {\sf\ftn (d)} of \Sref{obs} and \Eref{sc}
are fulfilled. Through Eq.~(\ref{ns}) we can then extract $n_{\rm
s}$ and $\alpha_{\rm s}$. Following this strategy, in
\sFref{fig5}{a} [\sFref{fig5}{b}] we display the allowed values of
$\ns$ [$M$] versus $\ck$ with $m\leq10^6~\TeV$, $\kappa=10^{-5}$
and $\lambda=\kp$ (solid lines) $\lambda=5\kp$ (dashed lines) and
$\lambda=0.5\kp$ (dotted lines). The region of \sEref{nswmap}{a}
is also limited by thin lines. We observe that as $\ck$ increases,
$\ns$ decreases entering the observationally favored region of
\sEref{nswmap}{a}. On the other hand, $M$ increases with $\ck$
until a certain $\ck\simeq0.03-0.05$ and then decreases.
Surprisingly the value of $\ck$, at which the maximum $M$ is
encountered, corresponds more or less to the central observational
$\ns\simeq0.96$. We also observe that increasing $\ld$ above $\kp$
with fixed $\ck$, $\ns$ drops but $M$ raises. These results can be
understood as follows: As $\ld/\kp$ elevates $\sgc$ decreases --
see \Eref{sc} -- and therefore, $\sg_*$ decreases, with fixed
$\widehat{N}_*$. This effect causes an increase of
$|f_{,\sg}(\sg_*)|$ and $f_{,\sg\sg}(\sg_*)$ -- see \Eref{nmhi1}.
As a consequence, $M$ increases too, since $M$ is proportional to
$f_{,\sg}^{1/2}$ due to \Eref{Prob}. Also, $|\eta|$ increases --
according to \Eref{nmhi3} -- and so, $\ns$ drops efficiently --
see \Eref{ns}.

\begin{figure}[!t]\vspace*{-.16in}
\hspace*{-.22in}
\begin{minipage}{8in}
\epsfig{file=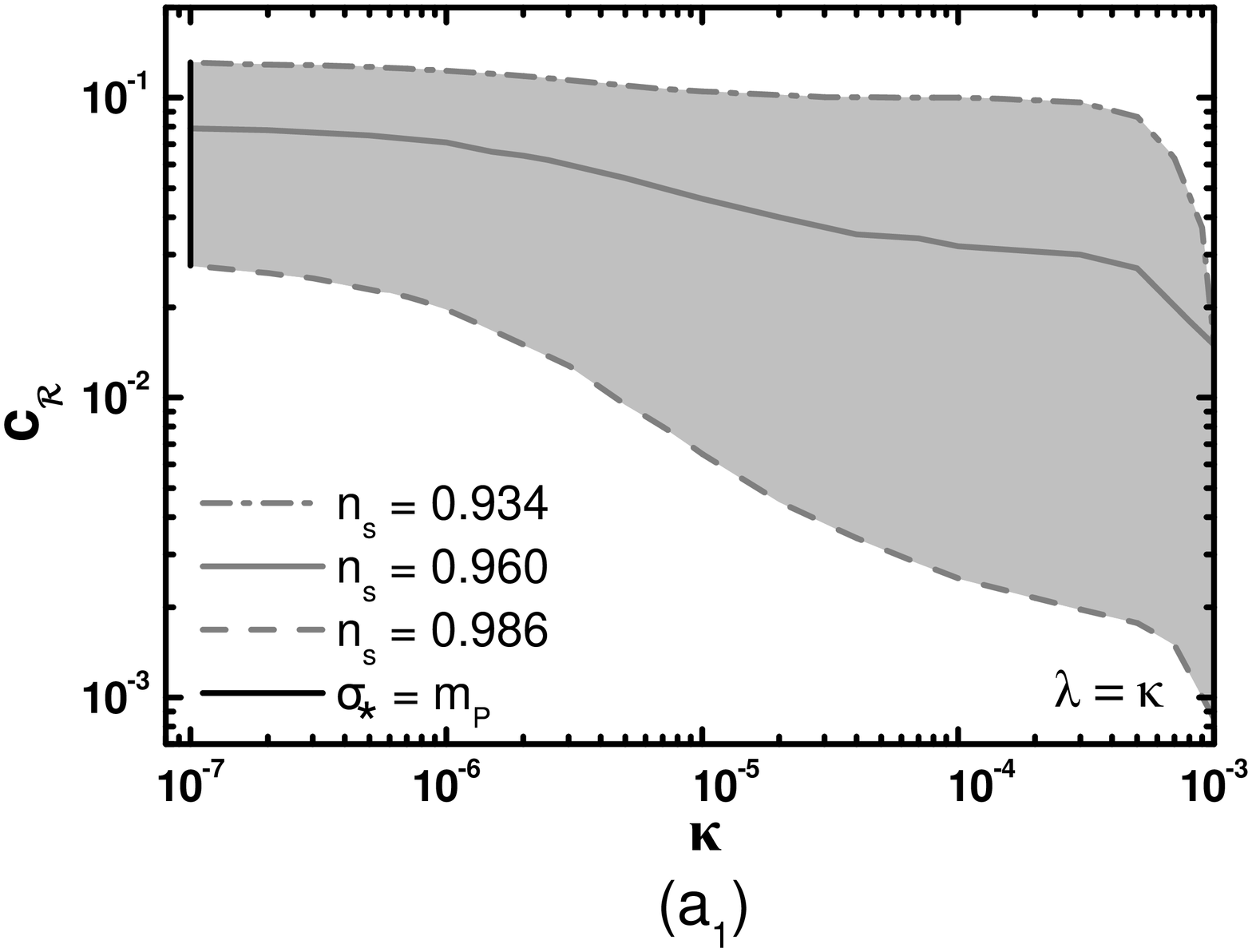,height=3.6in,angle=-90}
\hspace*{-1.2cm}
\epsfig{file=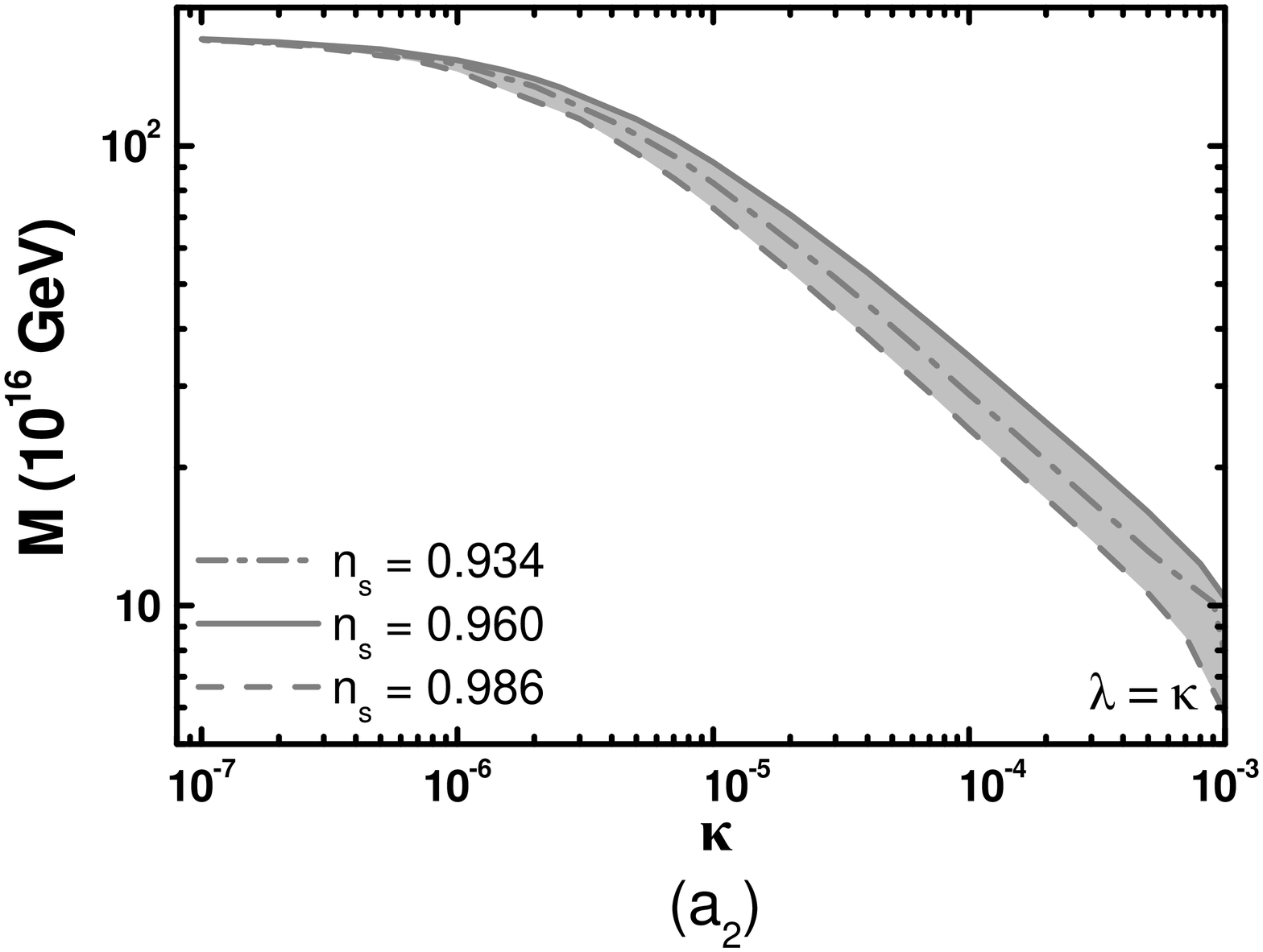,height=3.6in,angle=-90} \hfill
\end{minipage}\vspace*{-.1in}
\hfill \hspace*{-.22in}
\begin{minipage}{8in}
\epsfig{file=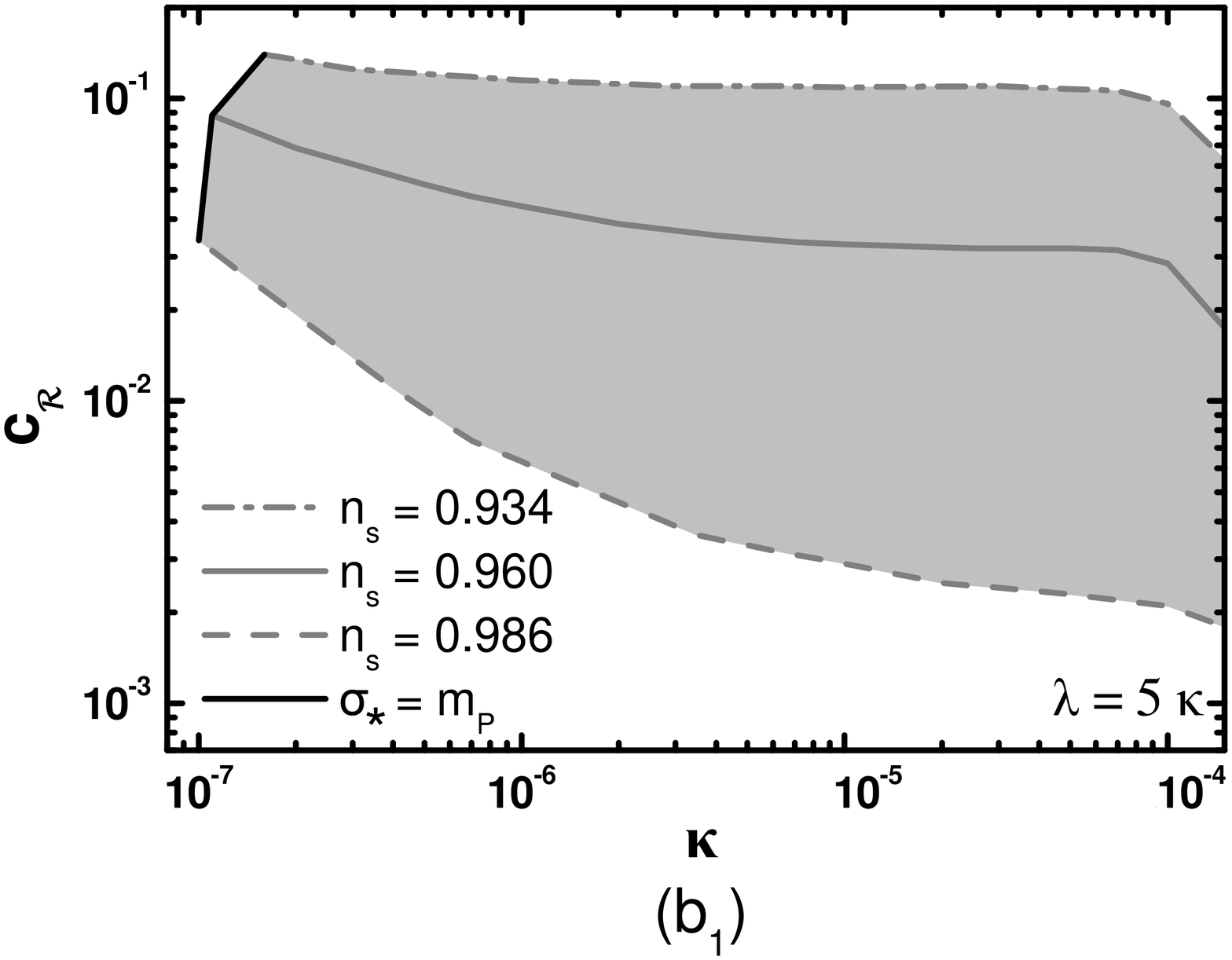,height=3.6in,angle=-90}
\hspace*{-1.2 cm}
\epsfig{file=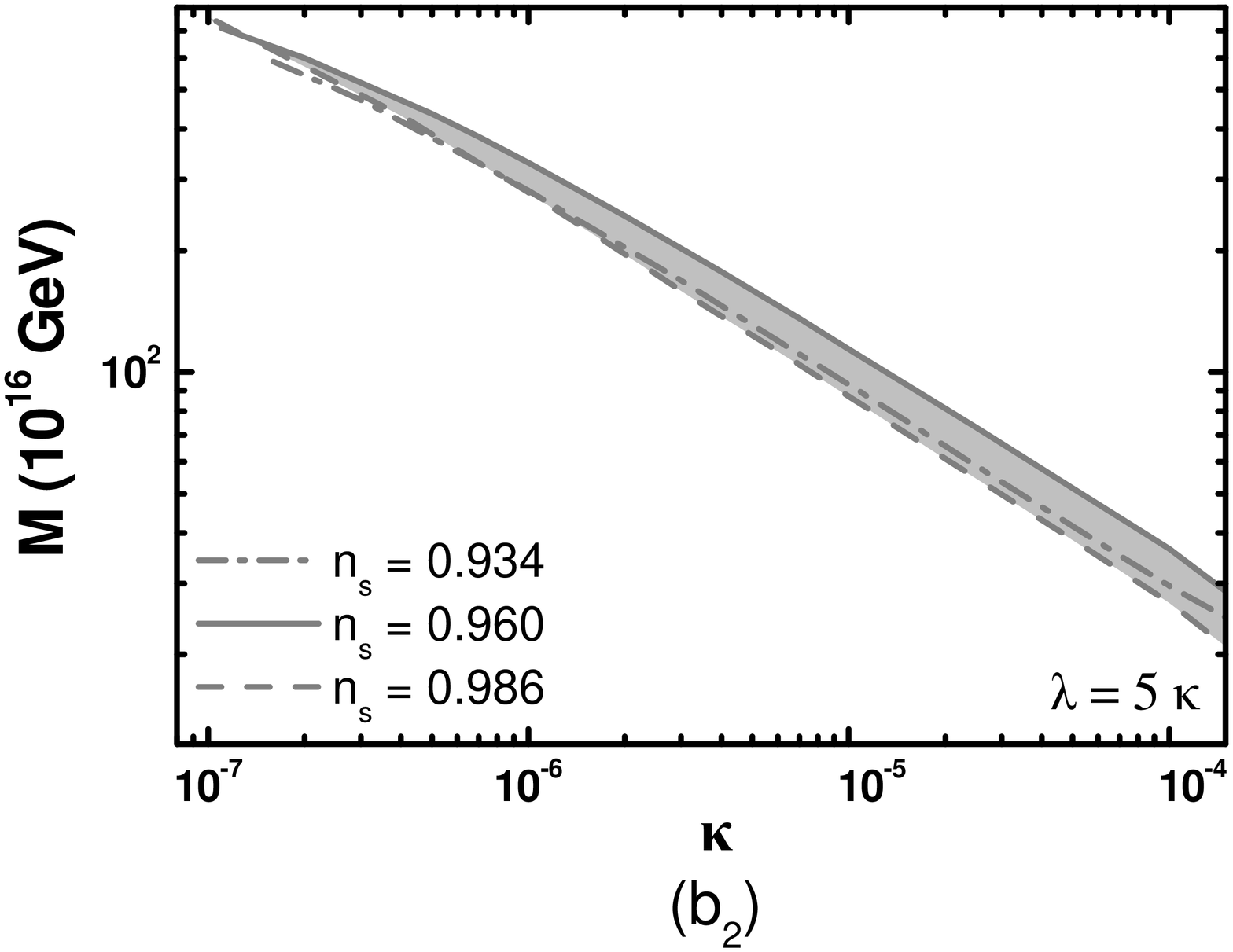,height=3.6in,angle=-90} \hfill
\end{minipage}\vspace*{-.1in}
\hfill \hspace*{-.22in}
\begin{minipage}{8in}
\epsfig{file=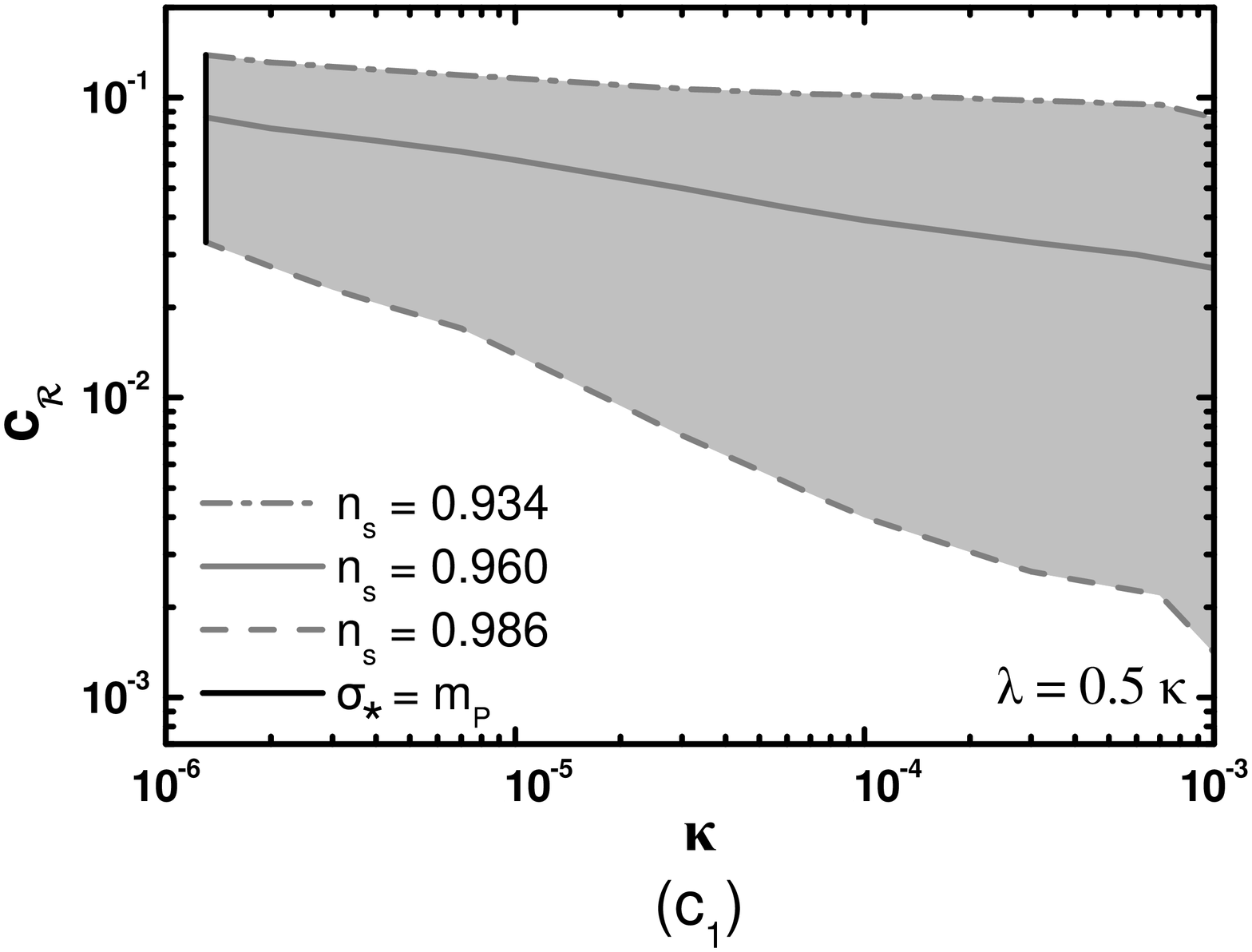,height=3.6in,angle=-90}
\hspace*{-1.2cm}
\epsfig{file=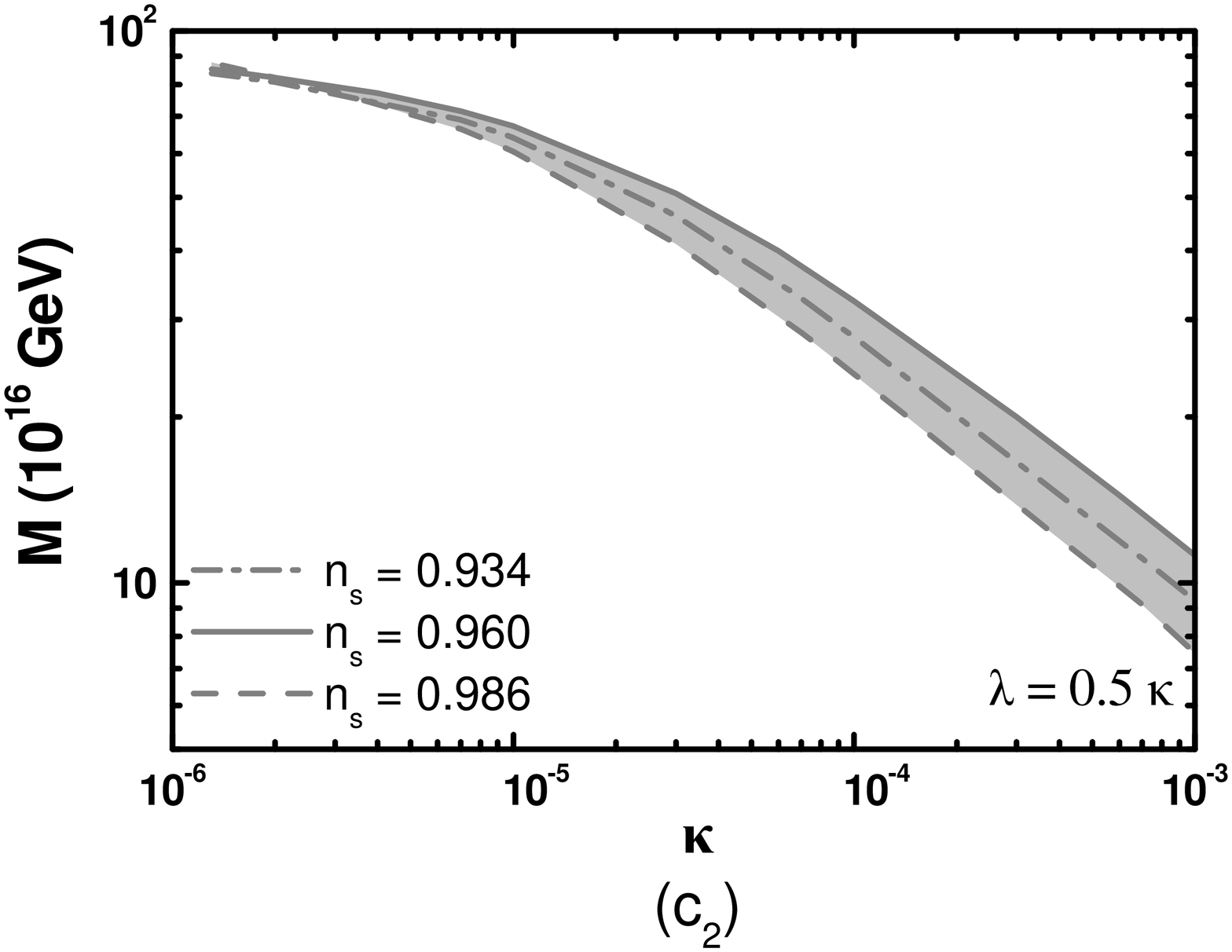,height=3.6in,angle=-90} \hfill
\end{minipage}%\vspace*{-.155in}
\hfill\begin{center}\renewcommand{\arraystretch}{1.1}
\begin{tabular}{|c|lll|l|ll|}\hline
Fig.&$\;\;\;\;\kappa/10^{-3}$&$\;\ck/10^{-2}$&$M/10^{16}~\GeV$
&$\Trh/10^{8}~\GeV$&$\;\;\;\;\;\Dcex$&$\Dex/10^{-2}$\\\hline\hline
{\ftn \sf (a$_1$), (a$_2$)}&
$0.0002-1$&$7.8-1.5$&$\;\;\;168-10$&$0.0028-3.5$&$0.016-7$&$0.91-31$\\
{\ftn \sf (b$_1$), (b$_2$)}&$0.00011-0.2$&$8.8-1.8$&$\;\;\;711-25$&$0.12-13$&$0.21-18$&$9.6-32$\\
{\ftn \sf (c$_1$), (c$_2$)}&$0.0013-1$&$8.6-2.7$&$\;\;\;85-11$&$0.006-2.1$&$0.03-4$&$1.5-32$\\
\hline
\end{tabular}
\end{center}\vspace*{-.08in}
\hfill \vchcaption[]{\sl\small Allowed (lightly gray shaded) by
the restrictions of \Sref{obs} consistently with \Eref{sc} areas
in the $\kappa-\ck$ [$\kappa-M$] plane ({\ssz \sf a$_1$}, {\sf\ssz
b$_1$} and {\sf\ssz c$_1$}) [({\ssz \sf a$_2$} {\sf\ssz b$_2$} and
{\sf\ssz c$_2$})] for non-MHI. We take $\kappa=\lambda$ and
$m\leq10^6~\TeV$ ({\ssz \sf a$_1$} and {\sf\ssz  a$_2$}) or
$m=10^{8}~\TeV$ and $\lambda=5\kappa$ ({\ssz \sf b$_1$} and
{\sf\ssz  b$_2$}) or $m=10^{7}~\TeV$ and $\lambda=0.5\kappa$
({\ssz \sf c$_1$} and {\sf\ssz  c$_2$}). The conventions adopted
for the various lines are also shown. The allowed ranges of the
various parameters for $\ns=0.96$ are listed in the
table.}\label{fig4}
\end{figure}

Confronting non-MHI with all the constraints of \Sref{obs}
consistently with \Eref{sc}, we can delineate the allowed (lightly
gray shaded) regions in the $\kappa-\ck$ [$\kappa-M$] plane as in
\Fref{fig4}-{\ftn\sf (a$_1$), (b$_1$)} and {\ftn\sf (c$_1$)}
[\Fref{fig4}-{\ftn\sf (a$_2$), (b$_2$)} and {\ftn\sf (c$_2$)}]. In
\Fref{fig4}-{\ftn\sf (a$_1$)} and {\ftn\sf  (a$_2$)} we take
$\ld=\kappa$. Our results for this choice are $m$-independent for
any $\kappa$ and $m\leq10^6~\TeV$. On the other hand, in
\Fref{fig4}-{\ftn\sf (b$_1$)} and {\ftn\sf (b$_2$)}
[\Fref{fig4}-{\ftn\sf (c$_1$)} and {\ftn\sf (c$_2$)}] we set
$\ld=5 \kappa$ and $m=10^8~\TeV$ and [$\ld=0.5 \kappa$ and
$m=10^7~\TeV$]. The conventions adopted for the various lines are
also shown in the left-hand side of each graph. In particular, the
gray dot-dashed [dashed] lines correspond to $n_{\rm s}=0.986$
[$n_{\rm s}=0.934$], whereas the gray solid lines have been
obtained by fixing $n_{\rm s}=0.96$ -- see Eq.~(\ref{nswmap}). For
$\kp$'s below the solid black line, our initial requirement in
\sEref{subP}{b} is violated. For $\kappa$'s larger than those
depicted in the graphs we do not find solutions consistent with
the imposed restrictions of \Sref{obs}. The upper bounds of the
allowed regions in the $\kappa-M$ plane come from $\ck$ leading to
$\ns=0.96$ -- see \sFref{fig5}{b}. Although this result may not
rigorously correct, it is accurate enough for our pictorial
purposes. In all cases, the allowed ranges of $\kp$'s -- although
restricted to values lower than $0.001$ -- are much more wide and
natural than the ones obtained for MHI -- c.f. Table of
\Fref{fig2}. Confining $\ns$ to its central observational value,
we obtain the ranges of the various parameters arranged in the
Table of \Fref{fig4}. We observe there that, for fixed $\ns$ and
increasing $\kappa$, $\ck$ and $M$ decrease whereas $\Dcex$ and
$\Dex$ increase. As a consequence, for any $m$, the tuning
regarding $\Dcex$ is greatly alleviated compared to the outputs of
MHI, whereas we are let with the usual mild tuning required for
$\Dex$. This is present to any inflationary hilltop model -- c.f.
\cref{gpp}. The allowed $M$'s mostly exceed the SUSY grand
unification scale, $M_{\rm GUT}\simeq2.86\cdot10^{16}~{\rm GeV}$,
whereas $\Trh$ mostly increases with $\kp$, as can be noticed via
\eqs{GTrh}{nmhi1}.

From our findings, we can conclude that: {\sf\ftn (i)} the
required $\ck$'s are rather low and so, complications related to
the hierarchy between the inflationary scale and the effective
cutoff of the theory are avoided; {\sf\ftn (ii)} our results
depend rather weakly on the variation of $m$, for $m\leq5\cdot
10^8~\TeV$; {\sf\ftn (iii)} as $m$ raises above $5\cdot 10^8~\TeV$
and $\kp$ drops below $0.001$, \sEref{subP}{b} is eventually
violated and so, our scheme becomes unapplicable; {\sf\ftn (iv)}
similarly to MHI, $\as$ and $r$ turn out to be negligibly small.

\addtolength{\textheight}{1.cm}
\newpage
%%%%%%%%%%%%%%%%%%%%%%%%%%%%%%%%%%%%%%%%%%%%%%%%%%%%%%%%%%%%%%%%%%%%%
\renewcommand{\arraystretch}{1.}

%%%%%%%%%%%%%%%%%%%%%%%%%%%%%%%%%%%%%%%%%
\addtolength{\textheight}{-1.cm}

As in the case of non-MCI, our proposal remains intact even if we
add fermion-dominated one-loop radiative corrections in \Eref{Vhi}
-- c.f. \cref{hirc} -- provided the values of the relevant Yukawa
coupling constant, $h$, remains lower than about $10^{-4}$. For
$h$'s close to this value, the decay width of the inflaton, due to
this channel dominates over the one given by \Eref{GTrh}.

\section{Conclusions}\label{con}

We considered the non-SUSY version of CI (driven by quadratic
potential) and HI, assuming a non-minimal coupling function,
$f(\sg)$, between  the inflaton field and the Ricci scalar
curvature. Using the freedom of choosing this scalar function, we
deliberated CI from the problem of \trns\ inflaton values and
achieved observationally acceptable $\ns$'s for a wide range of
the parameters of HI. As a bonus, the selected $f(\sg)$'s give
rise to Yukawa-type interactions between the inflaton and matter
fields leading to a successful post-inflationary reheating.
Afterwards, the proposed $f(\sg)$'s reduce to unity and so, the
Einstein gravity is naturally recovered.

Specifically, the adopted forms of $f(\sg)$ are given by
\Eref{fci} and \Eref{fhi} for non-MCI and non-MHI, respectively.
In both cases, the parameter $\ck$ involved in $f(\sg)$ can be
constrained so as the results of the inflationary models can be
reconciled with a number of theoretical and observational
restrictions. Our results are as follows:

\begin{itemize}

\item In the case of non-MCI, we find
$625\lesssim\ck\lesssim2.1\cdot10^7$ resulting to $\ns\simeq0.955$
and $r\simeq(0.2-0.22)$ for $n=+1$ and $83\lesssim\ck\lesssim3120$
resulting to $\ns\simeq0.967$ and $r\simeq(0.002-0.003)$ for
$n=-1$. In sharp contrast to MCI, only \sub values of the inflaton
field in the Jordan frame are utilized avoiding, thereby,
destabilization of the inflationary scenarios from possible
corrections caused by quantum gravity. Comments on the naturalness
of the models are also given.

\item In the case of non-MHI, the chosen $f(\sg)$ leads to
hilltop-type inflation for a wide range for $\kp$'s. As a
consequence, observationally acceptable results require a
proximity between the values of the inflaton field at the maximum
of the potential and at the horizon crossing of the pivot scale.
The amount of this tuning was measured by the quantity $\Dex$
defined in \sEref{dms}{b}. E.g., for $m\leq10^6~\TeV$ and the
observationally central value of $\ns$, we find
$\ck\simeq(0.015-0.078)$ with $M\simeq(1-16.8)\cdot10^{17}~\GeV$,
$\lambda=\kappa\simeq(2\cdot 10^{-7}-0.001)$ and
$\Dex\simeq(0.91-32)\%$. Compared to MHI, we find that the
observational requirements can be satisfied without tuning
severely neither $\kp$ nor $\Dcex$ defined in \sEref{dms}{a} even
for low $m$'s -- see Tables of Figs.~\ref{fig2} and \ref{fig4}.
Therefore, the proposed non-MHI is more favored by the current
data.

\end{itemize}

We explicitly checked that, for both models of non-MI, the
proposed scheme remains valid even if an extra coupling of the
inflaton to fermions exists, provided that the relevant coupling
constant is somewhat suppressed. If these fermions are identified
with right-handed neutrinos, baryogenesis via non-thermal
leptogenesis \cite{inlept} is, in principle, possible -- in the
case of HI, baryogenesis can be also accomplished if only the
waterfall field is coupled to right-handed neutrinos. Note that,
in our framework, the decay of the inflaton to right-handed
neutrinos is also possible due to curvature-induced
\cite{reheating} couplings. However, the resulting decay width is
reduced \cite{reheating} compared to this given by \Eref{GTrh} and
so, the produced lepton asymmetry is lower than the expectations
for all possible masses of right-handed neutrinos. On the other
hand, since baryogenesis can be realized in a variety of ways --
see, e.g., \cref{kolb, baryo} -- we opted not to complicate our
presentation with secondary mechanisms which may or may not affect
the inflationary observable quantities.

It would be interesting to investigate if a similar realization of
non-MI can be accomplished in the framework of SUGRA, along the
lines of \cref{susy}. In such a case, the inflaton of non-MCI
could be identified with one of the right-handed sneutrinos. On
the other hand, a possible SUSY version \cite{susyhybrid, gpp} of
non-MHI could become compatible with larger (and more natural)
values of the relevant coupling constant $\kp=\ld$.

\end{document}